\documentclass[twocolumn]{aastex62}
\usepackage{bibentry,natbib,CJK}
\usepackage{graphicx}
\usepackage{soul}

\received{***}
\revised{***}
\accepted{***}

\submitjournal{ApJ}
\shortauthors{Xu et al.}

\begin{document}
\begin{CJK*}{UTF8}{gbsn}

\title{STAR FORMATION IN MASSIVE GALAXIES AT REDSHIFT $z \sim 0.5$}

\correspondingauthor{Yipeng Jing}
\email{ypjing@sjtu.edu.cn}
\correspondingauthor{Chengze Liu}
\email{czliu@sjtu.edu.cn}

\author{Kun Xu}
\affil{Department of Astronomy, School of Physics and Astronomy, Shanghai Jiao Tong University, Shanghai, 200240, China}

\author{Chengze Liu}
\affil{Department of Astronomy, School of Physics and Astronomy, Shanghai Jiao Tong University, Shanghai, 200240, China}

\author{Yipeng Jing}
\affil{Department of Astronomy, School of Physics and Astronomy, Shanghai Jiao Tong University, Shanghai, 200240, China}
\affil{Tsung-Dao Lee Institute, and Shanghai Key Laboratory for Particle Physics and Cosmology, Shanghai Jiao Tong University, Shanghai, 200240, China}

\author{Yunchong Wang}
\affil{Department of Astronomy and Center for Astrophysics, Tsinghua University, Beijing, 100084, China}
\affil{Kavli Institute for Particle Astrophysics and Cosmology, Department of Physics, Stanford University, Stanford, 94305, USA}

\author{Shengdong Lu}
\affil{National Astronomical Observatories, Chinese Academy of Sciences, 20A Datun Road, Chaoyang District, Beijing 100101, China\\}
\affil{University of Chinese Academy of Sciences, Beijing 100049, China\\}

\begin{abstract}
It is believed that massive galaxies have quenched their star formation because of active galactic nucleus feedback. However, recent studies have shown that some massive galaxies are still forming stars. We analyze the morphology of star formation regions for galaxies of stellar mass larger than 10$^{11.3}$ M$_{\odot}$ at around redshift $z_r=0.5$ using $u-z$ color images. We find that about $20\%$ of the massive galaxies are star-forming (SF) galaxies, and most of them ($\sim 85\%$) have asymmetric structures induced by recent mergers. Moreover, for these asymmetric galaxies, we find that the asymmetry of the SF regions becomes larger for bluer galaxies. Using the Illustris simulation, we can qualitatively reproduce the observed relation between asymmetry parameter and color. Furthermore, using the merger trees in the simulation, we find a correlation between the color of the main branch galaxies at $z_r=0.5$ and the sum of the Star Formation Rates (SFRs) of the recently accreted galaxies, which implies that star formation of the accreted galaxies has contributed to the observed star formation of the massive (host) galaxies ({\it ex situ star formation}).  Furthermore, we find two blue and symmetric galaxies, candidates for massive blue disks, in our observed sample, which indicates that about $\sim 10\%$ of massive SF galaxies are forming stars in the normal mode of disk star formation ({\it in situ star formation}).  With the simulation, we find that the disk galaxies at $z_r \approx 0.5$ should have experienced few major mergers during the last 4.3 Gyrs. 
\end{abstract}

\keywords{galaxy: merger---galaxy: star formation---galaxy: massive galaxy}

\section{Introduction}
\label{sec:intro}

It has been found with observations and numerical simulations that massive galaxies are mostly early-type galaxies~\citep[ETGs, e.g.,][]{2013MNRAS.432.1862C}. Because of their environments, massive galaxies are believed to form via mergers that can destroy their disk structures~\citep[e.g.][]{2010MNRAS.403.1009M,2009ApJ...706L.173B}. As a result, compared to late-type galaxies (LTGs), ETGs have smaller disk components and are mainly dominated by bulges. Furthermore, many studies have found that the mass of the central supermassive black hole is tightly correlated with the bulge stellar mass \citep{2004ApJ...604L..89H,2013ARA&A..51..511K,2000ApJ...539L...9F}. Thus, each massive galaxy is expected to have a super massive black hole (SMBH) at its galactic center.\par

In recent years, many studies \citep{1998A&A...331L...1S,2000ApJ...539..517B,2006ApJ...648..820K,2006MNRAS.367..864C,2016MNRAS.463.1431K} have found that various feedback processes are crucial for galaxy formation. Simulations without feedback produce galaxies too massive to match observations~\citep{2010ApJ...725.2312O}. The stellar-to-halo mass relation (SHMR) also indicates that galaxy formation efficiency is different at different halo mass which may be caused by different feedback processes~\citep{2009ApJ...695..900Y,2019MNRAS.488.3143B}. Due to the existence of SMBH, AGN (active galactic nucleus) feedback becomes dominating at the high halo mass end~\citep{2006ApJ...648..820K,2006MNRAS.367..864C}, which is so strong in massive galaxies that can quench the entire galaxy~\citep{2005ApJ...620L..79S,2008MNRAS.387...13K}.\par

Therefore, massive galaxies should be red in color and have stopped their SF, typical of ETGs. However, recent studies have found that there are still ongoing star formation activities in some of the most massive galaxies~\citep{2007ApJS..173..619K,2012MNRAS.421..314C}. Studies focusing on the circumgalactic medium (CGM) also confirmed that cold gas still exists in the CGM of red massive galaxies~\citep{2012MNRAS.422.1835S,2018A&A...618A.126O,2019ApJ...883....5B}, which provides the ingredients for SF. \par

Beside feedback processes, galaxy mergers also play an important role in galaxy formation models. Previous works have noticed that the traditional monolithic collapse model (\citealt{1962ApJ...136..748E}, \citealt{1969MNRAS.145..405L,2013ARA&A..51..393C}) could not explain the evolution of ETGs, especially for the massive ETGs. Thus, models involving galaxy mergers, like the two-phase galaxy formation model~\citep{2010ApJ...725.2312O,2012ApJ...754..115J,2013ApJ...766...71R,2016MNRAS.458.2371R} have been proposed. Simulations have also shown that the growth of ETGs is mainly through the accretion of stars from other galaxies rather than \textit{in situ} SF~\citep{2010ApJ...709.1018V,2011ApJ...739L..44D}. \par

In this paper, we will study the morphology of star forming regions in massive galaxies with stellar mass larger than  $2 \times 10^{11}$ M$_{\odot}$ at redshift $z_r \approx 0.5$~\footnote{In order to distinguish with the label for the $z$ band photometry, we use $z_r$ for redshift}. If star formation in the massive galaxies is mainly from merged young (smaller) galaxies (i.e., {\it ex situ}), the star formation regions are expected to be irregular with merging signatures. In contrast, if the surrounding gas is cooling and forming stars around a massive galaxy (i.e., {\it in situ}), a regular star forming region is expected. Therefore, the morphologies of star forming regions could be an effective probe to quantify the modes of star formation in massive galaxies.

The rest frame color $NUV-r$ is known to be a good indicator for the specific star formation rate (sSFR) \citep{2005ApJ...629L..29B}. At the redshift of our interest (i.e., $z_r \sim 0.5$), the observed $u-z$ color index is a good proxy for the rest-frame $NUV-r$. Since star formation is generally weak in massive galaxies, the deep, high quality images in $u$ and $z$ bands are essential for the SFR estimation and the visual inspection of the morphologies of SF regions. The combination of CFHT Large Area U-band Deep Survey \citep[CLAUDS, $u$-band,][] {2019MNRAS.489.5202S} and Hyper Suprime-Cam Subaru Strategic Program \citep[HSC-SSP; $z$-band,][]{2018PASJ...70S...8A} gives us a unique opportunity to carry out such a study. We also compare our observational results with the Illustris Simulations \citep{2014MNRAS.444.1518V,2014Natur.509..177V,2014MNRAS.445..175G,2015A&C....13...12N,2015MNRAS.452..575S} to interpret the origin of star formation in massive galaxies at $z_r \approx 0.5$.

This paper is organized as follows. In Section \ref{sec:data}, we introduce the data selection process in both the observation and the simulation. In Section \ref{sec:observation} we analyze the morphology of star forming regions in the observation. And in Section \ref{sec:simulation}, the simulation data is studied and compared with the observation. Our conclusions are summarized in Section \ref{sec:conclusion}. Throughout this paper, we adopt the following cosmological parameters: $H_0 = 70.0$ km s$^{-1}$ Mpc$^{-1}$, $\Omega_M = 0.3$ and $\Omega_{\Lambda} = 0.7$. All the magnitudes we used are in the AB magnitude system \citep{1983ApJ...266..713O}.\par

\section{Data and sample selection}
\label{sec:data}

To select the massive galaxies with star forming regions at redshift $z_r \sim 0.5$, we use the data from a combination of several photometric and spectroscopic surveys. The redshift information is drawn from the CMASS sample of the Baryon Oscillation Spectroscopic Survey \citep[BOSS,][]{2012ApJS..203...21A, 2012AJ....144..144B}. The CMASS sample consists of massive galaxies ($i < 19.9$ mag) with the redshift range of $0.4 < z_r < 0.8$, which covers the redshift of our interest (i.e., $z_r \sim 0.5$).

We use the data from the VIMOS Public Extragalactic Redshift Survey (VIPERS) Multi-Lambda Survey \citep{2014A&A...566A.108G, 2016A&A...590A.102M} to estimate the stellar masses of candidate galaxies. They collected the ultraviolet bands ($FUV$ and $NUV$) data from GALEX survey \citep{2005ApJ...619L...1M}, the optical data ($u^*$, $g$, $r$, $i$ and $z$) from the Canada-France-Hawaii Telescope Legacy Survey\footnote{http://www.cfht.hawaii.edu/Science/CFHTLS/} (CFHTLS) and the near infrared $K_s$ band data that performed with the CFHT/WIRCam instrument \citep{2004SPIE.5492..978P}. \cite{2016A&A...590A.102M} combined all the data sets and built-up a homogeneous multi-bands ($FUV$, $NUV$, $u^*$,  $g$, $r$, $i$, $z$, and$K_s$) catalogue, which is optimal to measure the physical properties of galaxies, e.g., stellar masses \citep{2016A&A...590A.103M}.

Basically, we use the stellar population synthesis models of \cite{2003MNRAS.344.1000B} to  compute the physical properties of galaxies. In these calculations, the \cite{2003PASP..115..763C} Initial Mass Function (IMF) is adopted. We assume a delayed star formation history $\phi(t)\approx t \exp(-t/\tau)$ with $\tau$ taken from $10^7 yr$ to $1.258\times 10^{10} yr$ with an equal logarithmic interval $\Delta \lg \tau=0.1$. Three metallicities $Z/Z_\odot=0.4,1$ and $2.5$ are considered, where $Z_\odot$ is the metallicity of the Sun. We use the extinction law of \cite{2000ApJ...533..682C} with the dust reddening in the range $0<E(B-V)<0.5$. Finally we use the spectral energy distribution (SED) fitting code \textsc{Le PHARE} \citep{2002MNRAS.329..355A,2006A&A...457..841I} to get the physical properties, e.g., stellar mass and SFR.

We use the rest-frame $NUV-r$ color index as the star formation indicator \citep{2005ApJ...629L..29B}, which redshifts approximately to $u-z$ at redshift $z_r \sim 0.5$. To better estimate the star formation and check the morphologies of star forming regions, we use deeper $u$ band data from CLAUDS \citep{2019MNRAS.489.5202S} and deeper $z$ band data from HSC-SSP \citep{2018PASJ...70S...8A}. CLAUDS is a deep $u$ band imaging survey that uses MegaCam instruments \citep{2003SPIE.4841...72B} on the CFHT. CLAUDS covers four fields: XMM-LSS, E-COSMOS, ELAIS-N1 and DEEP2-3 with a median depth of $u=27.1$ mag (5 $\sigma$, see \citealt{2019MNRAS.489.5202S} for details). HSC-SSP is an ongoing multi-wavelength ($g$, $r$, $i$, $z$, $y$ and several narrow bands) imaging survey using HSC instrument \citep{2018PASJ...70S...1M} on Subaru telescope. The second HSC-SSP data release has been published \citep{2019PASJ...71..114A} with a $z$ band depth of $26.6$ mag in four deep fields, which are the same as CLAUDS fields. In this study, we focus on the XMM-LSS field ($3.5$ deg$^2$) which is an overlap region of VIPERS, BOSS/CMASS, CLAUDS and HSC-SSP Deep layer.

There are 372 CMASS galaxies in XMM-LSS field in the redshift range of $0.4 < z_r < 0.8$. As noticed by previous studies \citep[e.g.,][]{2013ApJ...767..122G}, the CMASS sample is complete only for massive galaxies with $M_*>10^{11.3}$ M$_{\odot}$. Furthermore, we have compared the stellar mass function of the CMASS galaxies with that of the VIPERS galaxies (Xu et al. in preparation), and have confirmed the completeness limit at $\sim 10^{11.3}$ M$_{\odot}$. Therefore, we adopt a stellar mass cut at $10^{11.3}$ M$_{\odot}$ in this study. There are $85$ galaxies left with $M_*> 10^{11.3}$ M$_{\odot}$.

The final step of sample selection is the visual inspection of CLAUDS $u$ band and HSC-SSP $z$ band images. Two galaxies are eliminated from the final sample. One galaxy is heavily contaminated by a nearby bright star, and the other exhibits a significant strong lensing effect. Our final sample consists of 83 massive galaxies ($M_*> 10^{11.3}$ M$_{\odot}$) at redshift $z_r \sim 0.5$.

For the simulation, we choose mock galaxies from the Illustris Simulations~\citep{2014MNRAS.444.1518V,2014Natur.509..177V,2014MNRAS.445..175G,2015A&C....13...12N,2015MNRAS.452..575S}. Illustris is a series of cosmological hydrodynamical simulations with a cubic box of side length 106.5 Mpc. It follows the evolution of dark matter, stars, gas and supermassive black holes from $z_r=127$ to $z_r=0$, and can match with observations broadly \citep{2014MNRAS.445..175G}. In this paper, we use mock galaxy data from the Illustris-1 simulation run which has the highest mass and spatial resolution in the simulation series, i.e. mass resolutions of $m_{\mathrm{dm}} = 6.3\times 10^{6}\, \mathrm{M_{\odot}}$ and $m_{\mathrm{baryon}} = 1.3\times 10^6 \, \mathrm{M_{\odot}}$ for dark matter and baryons respectively. It has $1820^3$ resolution elements for both dark matter and gas, and adopts a gravitational softening length of 710 pc (note the softening length of the Voronoi gas cells is adaptive). To compare with observation, we define the stellar mass of each galaxy as twice the stellar mass within the half stellar mass radius, and select galaxies with stellar mass above 10$^{11.3}$M$_{\odot}$ at $z_r=0.5$ (snapshot 103). With these criteria, we have 81 mock galaxies in our Illustris sample. Moreover, we will use the \textsc{sublink}~\citep{2015MNRAS.449...49R} merger trees to trace their merger histories, as well as their redshifted images~\citep{2015MNRAS.447.2753T} to compare with our observational results. 

\section{Properties of the Observed Sample}
\label{sec:observation}

In this section, we study the observational sample and analyze the relation between their morphologies in the $u-z$ maps and their global $u-z$ colors (i.e., sSFR). For most of the massive galaxies without star formation, we would expect they are symmetric and red in the color. If the galaxies are still forming stars from cold gas ({\it in situ}), star formation is expected to be symmetric and they are expected to be blue in color. If star formation is mainly caused by the merging of small galaxies with young stars and/or cold gas, we would expect an asymmetric color map and a bluer global color. We will first make a visual inspection of the color maps, and then quantify their morphologies. We will also use the S\'ersic profile to fit their luminosity distribution in the $z$ band (approximately $r$ band in the rest frame) in order to find if there are blue disk galaxies at the massive end.

\begin{figure}
\epsscale{1.10}
\plotone{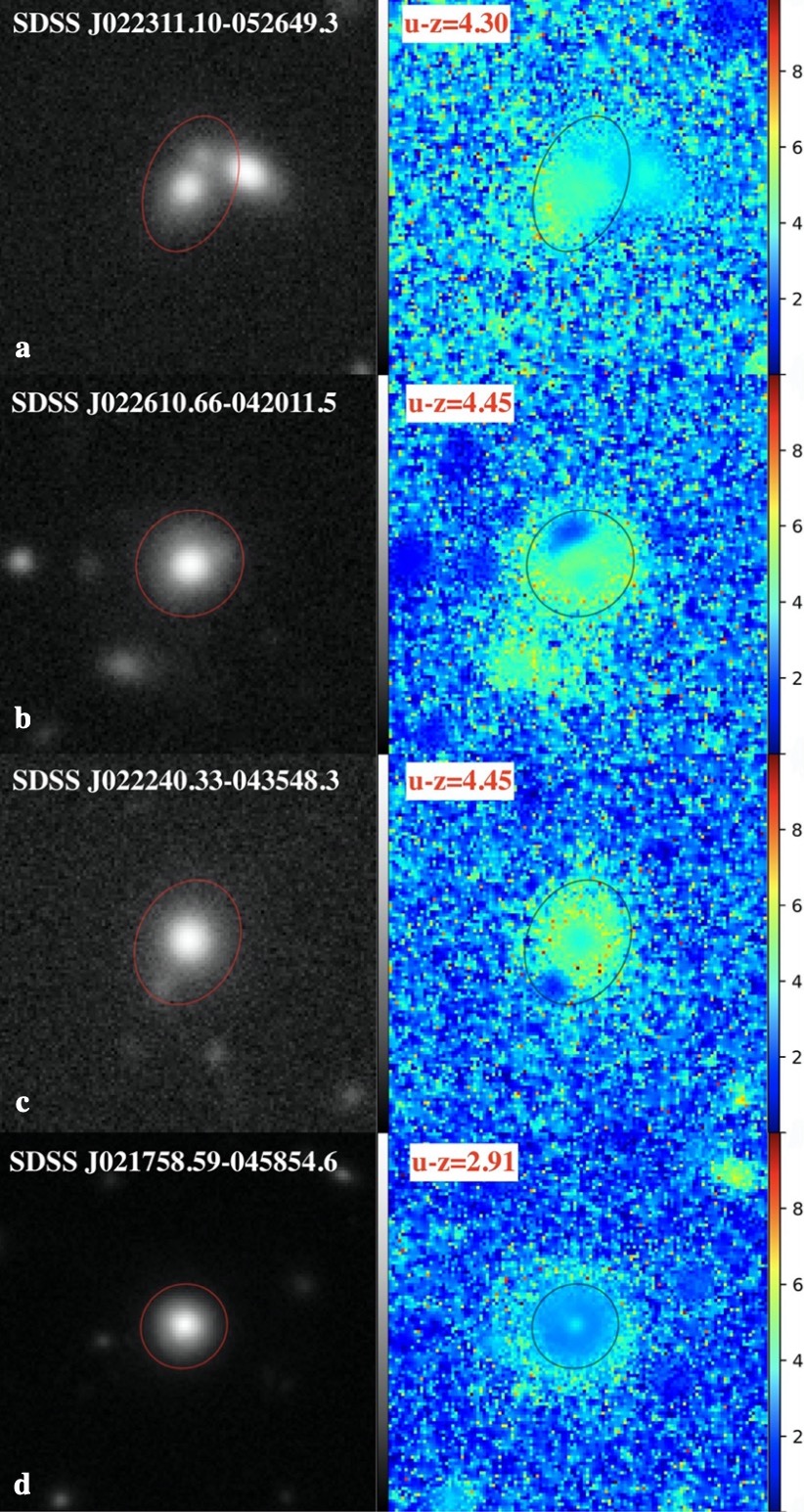}
\caption{The $z$ band images (left panels) and the $u-z$ color maps (right panels, the redder the coded color, the larger the color index) of four examples of SF galaxies in our sample. Each panel covers a $20\arcsec\times20\arcsec$ area centered on each galaxy. Red (left panels) and black (right panels) ellipses indicate the $z$ band Petrosian radii of the galaxies. The identifications and $u-z$ color indices of galaxies are given
in the upper left corners of the $z$ band images and $u-z$ color maps respectively. (a) Category I:  Asymmetric in both the star formation (i.e., $u-z$ color map) and the stellar distribution (i.e., $z$ band image). In this panel, the small galaxy (upper) is an infalling galaxy that trigger the star formation while the bigger one on the right of the main galaxy is a foreground galaxy according to the photo-z measurement ($zp=0.27\pm0.02$). (b) Category II: Asymmetric in the star formation, but symmetric in the stellar distribution. (c) Category I-II: Asymmetric in the star formation with a faint asymmetric structure in the stellar distribution. (d) Category III: Symmetric in both the star formation and the stellar distribution. \label{fig:f1}}
\end{figure}

\subsection{Visual Inspection}
\label{subsec:eyes}

First of all, we use the definition of \cite{2013MNRAS.436..697B} to classify SF and quenched galaxies. SF galaxies are defined to satisfy:
\begin{equation}
\label{eq:1}
\log{(\rm SFR)} > -0.6 + 0.65(\log{M_{\star}}-10).
\end{equation}
With this definition, there are $18$ SF galaxies in the observational sample, which is $\sim 22\%$ of the total $83$ galaxies. \par

All the $u-z$ color maps of the SF galaxies are generated using $u$ magnitude maps minus $z$ magnitude maps. We visually inspect the $z$ band images and $u-z$ color maps of the 18 SF galaxies. They are firstly classified as asymmetric and symmetric (Category III) galaxies according to the $u-z$ maps. Then the asymmetric galaxies are further classified according to their $z$ band images (Category I, II, I-II). Figure \ref{fig:f1} shows the $z$ band images and $u-z$ color maps for the example galaxies in different category (three asymmetric categories and one symmetric category).\par

We only consider the structures within the $z$ band Petrosian radii \citep[$\eta$=0.2,][]{1976ApJ...209L...1P} of the galaxies, which are shown as red and black ellipses in Figure~\ref{fig:f1}. As a result, out of the 18 SF galaxies, we find that 4 galaxies show symmetric structure (Figure \ref{fig:f1}d) while the other 14 galaxies show asymmetric structures (Figure \ref{fig:f1} abc) in the $u-z$ color maps. Of these 14 asymmetric (in $u-z$ color maps) galaxies, four galaxies (Category I) show asymmetric structures in their stellar distributions ($z$ band images, see Figure \ref{fig:f1}a) as well. These structures are infalling galaxies that are going to merge with the main galaxies. The active SF regions are tied to the infalling galaxies. The SF of these massive galaxies is mainly contributed by the existing young stars of the merged galaxies ({\it ex-situ}), though some star formation could be triggered by the interaction. In contrast, six galaxies (Category II) show clear asymmetric structures in their star formation, but their stellar distributions do not show visible merger signatures and look well relaxed (Figure \ref{fig:f1}b). The rest 4 galaxies (Category I-II) are intermediate, showing faint counterparts in the stellar distributions (see  Figure \ref{fig:f1}c). The galaxies in Category II are likely to be at more advanced stages of the mergers compared with those in Category I, i.e. the mergers have completed, while the galaxies of Category I-II have nearly completed their merges. It is plausible that the mergers have played an important role in the star formation activities of these galaxies. We list the properties of these 18 SF galaxies in Table~\ref{tab:t1}. Their $z$ band images and $u-z$ maps are given in the APPENDIX.

\begin{deluxetable*}{ccccccccc}
\tablenum{1}
\tablecaption{Properties of 18 SF galaxies in our sample.}
\label{tab:t1}
\tablewidth{0pt}
\tablehead{
\colhead{ID}&\colhead{M$_\ast$}&\colhead{redshift}&\colhead{SFR}&\colhead{$A_{abs}$}&\colhead{n\tablenotemark{a}}&\colhead{$u-z$}&\colhead{Category}&\colhead{$EW$([OII]+O[III])\tablenotemark{b}}\\
\colhead{}&\colhead({M$_{\odot}$)}&\colhead{}&\colhead{(M$_{\odot}\,$yr$^{-1}$)}&\colhead{}&\colhead{}&\colhead{(mag)}&\colhead{}&\colhead{\AA}
}
\startdata
SDSS J021922.97-051151.3 & $10^{11.39}$ & 0.46 & $10^{0.4}$ & 0.18 & \nodata\      & 5.4 & I    &  2.5 \\
SDSS J022301.82-040853.2 & $10^{11.38}$ & 0.63 & $10^{0.6}$ & 0.19 & \nodata\      & 3.0 & I    &  4.8 \\
SDSS J022612.55-044504.8 & $10^{11.39}$ & 0.68 & $10^{2.0}$ & 0.07 & \nodata\      & 2.5 & I    & 80.8 \\
SDSS J022311.10-052649.3 & $10^{11.35}$ & 0.50 & $10^{0.3}$ & 0.14 & \nodata\      & 4.3 & I    &  1.0 \\
SDSS J021604.25-045855.3 & $10^{11.46}$ & 0.50 & $10^{0.4}$ & 0.15 & \nodata\      & 3.7 & II   &  4.2 \\
SDSS J021654.87-042126.5 & $10^{11.30}$ & 0.44 & $10^{0.3}$ & 0.07 & \nodata\      & 4.2 & II   &  8.4 \\
SDSS J022429.47-045840.3 & $10^{11.59}$ & 0.49 & $10^{1.6}$ & 0.11 & \nodata\      & 5.4 & II   &  0.0 \\
SDSS J022610.66-042011.5 & $10^{11.47}$ & 0.50 & $10^{0.4}$ & 0.11 & \nodata\      & 4.4 & II   &  0.0 \\
SDSS J022356.95-051922.9 & $10^{11.30}$ & 0.62 & $10^{0.6}$ & 0.09 & \nodata\      & 5.6 & II   &  3.5 \\
SDSS J022255.51-045316.7 & $10^{11.37}$ & 0.60 & $10^{2.0}$ & 0.11 & \nodata\      & 3.8 & II   &  8.0 \\
SDSS J022533.32-053204.6 & $10^{11.60}$ & 0.57 & $10^{0.6}$ & 0.14 & \nodata\      & 6.1 & I-II &  6.5 \\
SDSS J022240.33-043548.3 & $10^{11.38}$ & 0.60 & $10^{0.4}$ & 0.10 & \nodata\      & 4.5 & I-II &  5.4 \\
SDSS J022633.21-042859.8 & $10^{11.46}$ & 0.63 & $10^{0.4}$ & 0.07 & \nodata\      & 4.5 & I-II &  5.7 \\
SDSS J022152.03-041201.1 & $10^{11.82}$ & 0.61 & $10^{1.9}$ & 0.11 & \nodata\      & 5.4 & I-II &  3.2 \\
SDSS J022343.67-035630.8 & $10^{11.49}$ & 0.56 & $10^{0.7}$ & 0.04 & $3.04\pm0.09$ & 3.7 & III  &  8.6 \\
SDSS J022322.00-045738.5 & $10^{11.44}$ & 0.78 & $10^{2.5}$ & 0.08 & $1.61\pm0.07$ & 3.0 & III  & 45.8 \\
SDSS J022644.62-040808.4 & $10^{11.35}$ & 0.71 & $10^{2.2}$ & 0.09 & $1.17\pm0.04$ & 3.5 & III  & 78.3 \\
SDSS J021758.59-045854.6 & $10^{11.34}$ & 0.63 & $10^{2.5}$ & 0.04 & $2.32\pm0.05$ & 2.9 & III  &  6.5 \\  
\enddata
\tablenotetext{a}{S\'ersic analysis is only done for symmetric galaxies since it's meaningless to perform it in merger systems.}
\tablenotetext{b}{The total equivalent width ($EW$) for emission line [OII]3727, [OIII]4959 and [OIII]5007. The $EW$ values for emission lines are drawn from public released SDSS data \citep{2015ApJS..219...12A}. Please note that the SDSS spectroscopic fiber diameter is 3 arcsec. So the SDSS spectra can only detect the star formations in the galactic centers.}
\end{deluxetable*}

Since we do not have spectroscopic redshift measurement for the infalling galaxies, they might be foreground or background projections. We can estimate the probability that such a projection happens. The typical luminosity of our selected galaxies is $m_z \approx 19$,  and let's assume that the infall galaxies are less than 2 magnitudes fainter. Furthermore, the angular separation between the pair of the galaxies is less than $3\arcsec$. The mean sky density of galaxies is calculated using the photometric catalog of the second public data release (PDR-2) of VIPERS~\citep{2014A&A...566A.108G} in W1 field. There are 195156 sources with $m_z<21$ in the total $72\ deg^2$ field. Then the probability that a galaxy of $z<21$ aligns randomly with one of our selected galaxies within $3\arcsec$  is $0.59\%$. Therefore the chance of a projection is quite small. An alternative approach is the photo-z measurements. However, the infalling galaxies are usually faint and close to the main galaxies, which will cause larger uncertainty in photometry and photo-z measurements. Fortunately, the infalling galaxies in Category I are not too close to the main galaxies. So we can check whether these infalling galaxies are projections or not. There are four Category I galaxies as listed in Table \ref{tab:t1}. The corresponding spectroscopy redshifts ($zs$) for main galaxies and photo-z ($zp$) values for infalling galaxies are:\\
 SDSS J021922.97-051151.3,\\
 $zs_{main} = 0.46$,~$zp_{infall} = 0.44 \pm 0.17 $;\\
 SDSS J022301.82-040853.2,\\
 $zs_{main} = 0.63$,~$zp_{infall} = 0.82 \pm 0.39 $;\\
 SDSS J022612.55-044504.8,\\
 $zs_{main} = 0.68$,~$zp_{infall} = 0.66 \pm 0.04 $;\\
 and SDSS J022311.10-052649.3,\\
 $zs_{main} = 0.50$,~$zp_{infall} = 0.46 \pm 0.04$.\\
As we can see, the redshift differences between main and infalling galaxies are within the errors. The galaxy SDSS J022301.82-040853.2 has the largest redshift difference, but the photo-z error of the infalling galaxy is huge ($0.82 \pm 0.39$). Basically, the photo-z measurements support our conclusion that the chance of the projection is low.\par

\begin{figure*}
\epsscale{1.0}
\plotone{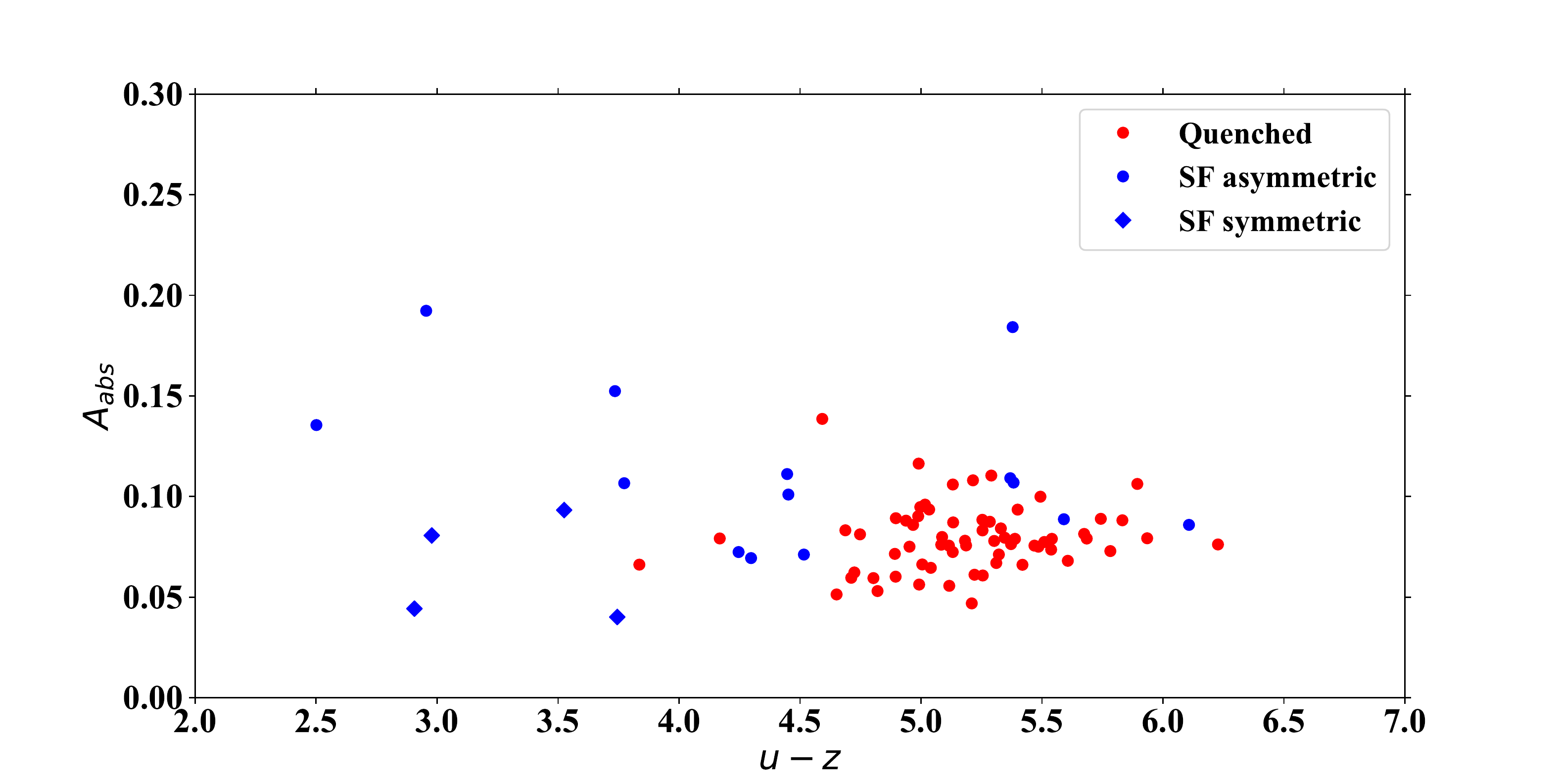}
\caption{$A_{abs}$ versus $u-z$ for the observational sample. Red and blue dots distinguish quenched from star-forming galaxies using Equation \ref{eq:1}. Star forming galaxies are sub-divided between symmetric (filled diamonds) and asymmetric (filed circle) based on the visual classifications described in Section \ref{subsec:eyes}. \label{fig:f2}}
\end{figure*}

Four galaxies in Category III do not show significant asymmetric signatures within the Petrosian radii either in their SF regions or in their stellar distributions (see Figure \ref{fig:f1}d).  It is worth noting that one of them (SDSS J022343.67-035630.8) has a close neighbor outside its Petrosian radius, which may have influences on its star formation. The other three, however, show regular, symmetric and vigorous star formation without any merger signatures. It is possible that these galaxies are in the regular mode of disk star formation ({\it in situ}). We will study their structures in more detail in the next subsections. One of them (SDSS J022322.00-045738.5) shows a small axis ratio ($b/a=0.67$) in their SF and stellar distribution morphologies, which may indicate that it has a large inclination angle.\par

To sum up, there are 18 SF galaxies in the observational sample according to our definition (Equation \ref{eq:1}). Observed visually, a majority of SF galaxies ($\sim 80\%$) seem to have been through recent mergers or to be undergoing mergers ({\it ex situ}). The rest of the SF galaxies do not display clear merger signatures within the Petrosian radii, and are likely in the regular mode of {\it in situ} disk star formation. \par

\subsection{Quantitative asymmetry}
\label{subsec:quantitative}

To make our results more objective and perform quantitative analysis on larger datasets in the future, we will quantify the asymmetry of the star formation regions of the observed galaxies. We use the definition of asymmetry suggested by \cite{2000ApJ...529..886C}:
\begin{equation}
\label{eq:2}
A_{abs} = \frac{\sum{\mid I_{0^{\circ}}-I_{180^{\circ}}\mid}}{2\sum{\mid I_{0^{\circ}}\mid}},
\end{equation}
where $I_{0^{\circ}}$ and $I_{180^{\circ}}$ are the intensity distributions in the original images and the images rotated by $180^{\circ}$. The sum is over all pixels within a prescribed matching region of the original and rotated images. Therefore, the completely symmetric images have $A_{abs}$ equal to 0, and the maximum theoretical value of $A_{abs}$ is 1 which describes the extreme asymmetric case. We calculate the $A_{abs}$ values for the galaxies based on their $u-z$ color maps and study the relation between their $A_{abs}$ and global $u-z$ colors. Details of the algorithm are as follows.\par

First we run \textsc{Sextractor}~\citep{1996A&AS..117..393B} in dual-image mode with the sources detected on the $z$ band (rest-frame $r$ band) image. We measure the central coordinates, Petrosian radii and automatic aperture magnitude (\textsc{MAG\_AUTO}, the most precise estimate of total magnitudes for galaxies) in both $u$ and $z$ bands. Then the $u-z$ maps are rotated around their centers by $180^{\circ}$ using the \textsc{IRAF}~\citep{tody1986iraf} rotation package, and the $A_{abs}$ value for each galaxy is calculated within the Petrosian radius. Finally, \textsc{MAG\_AUTO} is used to measure global $u-z$ color for each galaxy.

\begin{figure}
\epsscale{1.15}
\plotone{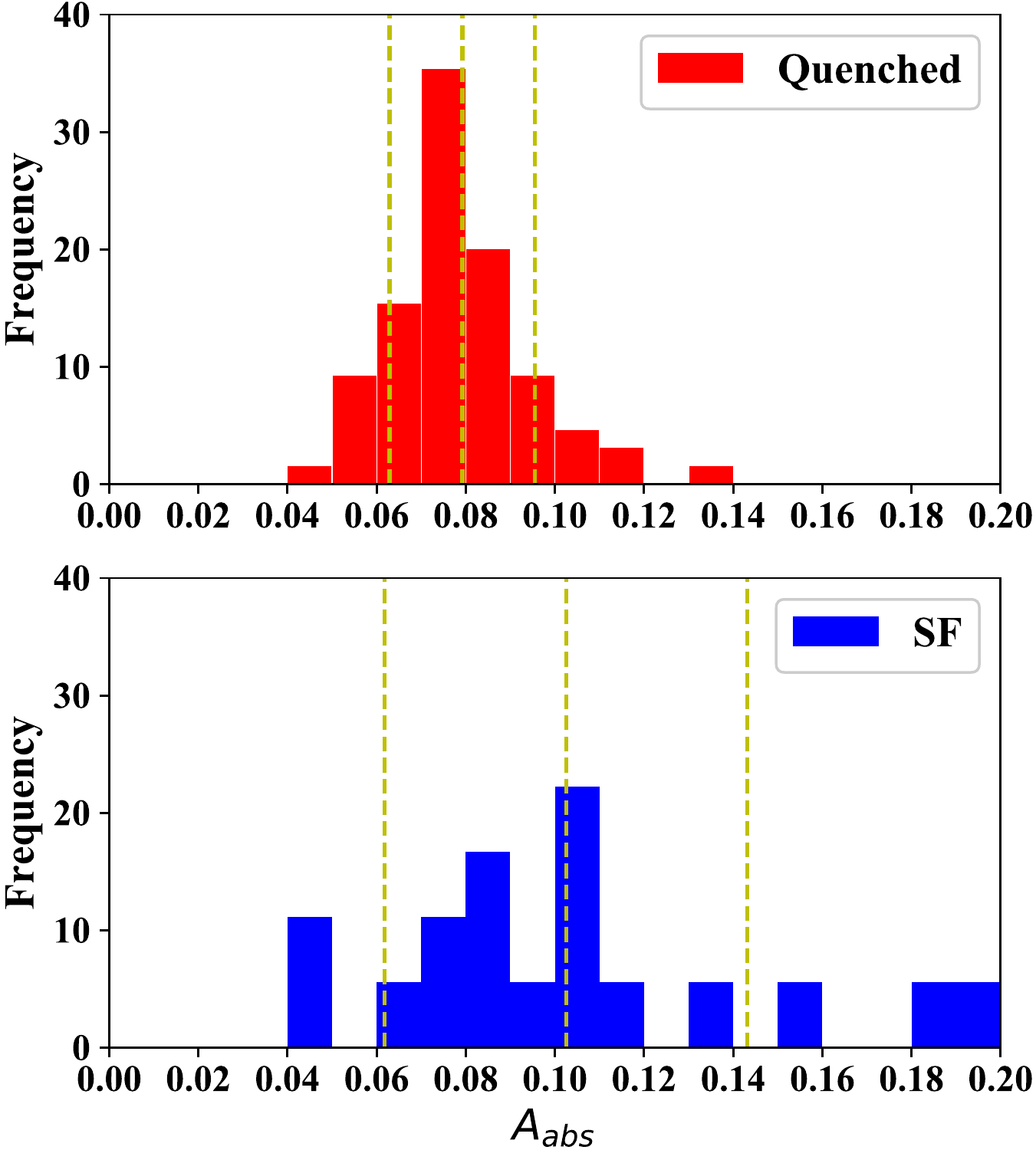}
\caption{Normalized distributions of $A_{abs}$ for SF (lower panel) and Quenched galaxies (upper panel). Mean values and 1$\sigma$ deviations are shown in each panel as dashed vertical lines. The KS-test \textit{p}-value of these two distribution is 0.0126.\label{fig:f3}}
\end{figure}

Figure \ref{fig:f2} shows the relation between the global $u-z$ colors and the asymmetry parameter $A_{abs}$ of galaxies in the observational sample. The colors of the dots indicate whether they are SF galaxies or quenched galaxies classified according to Equation~\ref{eq:1}. The four symmetric SF galaxies (Category III) are shown in blue diamonds. These four galaxies have $A_{abs} < 0.1$, consistent with our visual inspection. There seems to be a correlation between the color and the asymmetry as long as we ignore the four Category III galaxies. The bluer the galaxies, the higher the $A_{abs}$ parameters. Also, the SF galaxies tend to have higher $A_{abs}$ than the quenched ones at given color. To demonstrate their difference, we show the distribution of asymmetry parameter $A_{abs}$ in SF and quenched galaxies separately in Figure \ref{fig:f3}. The $A_{abs}$ value tends to be higher for SF galaxies. The difference in $A_{abs}$  between the SF and the quenched galaxies is statistically significant since the KS-test \textit{p}-value is only 0.0126. We note that the seeing effect in observation will generally reduce the asymmetry. In the meantime, because quenched galaxies have little star formation and they are faint in the $u$ band, the Poisson noise in the $u$ band images would lead to some level of asymmetry in quenched galaxies. Therefore, the intrinsic difference in the asymmetry between the SF and quenched galaxies could be larger. From the Figure \ref{fig:f2} and \ref{fig:f3}, we can see that the signatures of the post or ongoing mergers occurred in most of the SF galaxies, which indicates that the SF activities in the massive galaxies may be mainly caused by mergers. We will discuss these points in the section of simulation analysis. \par

\subsection{S\'ersic analysis} 
\label{subsec: Sersic}

It is interesting to quantify the morphology of the stellar component to find out whether the massive symmetric SF galaxies in Category III are disks or spheroids.  We use the S\'ersic profile to analyze the stellar morphology of these galaxies. \par

The S\'ersic indices of the galaxies are calculated by fitting the $z$ band luminosity profiles of the galaxies~\citep{2005PASA...22..118G} using:
\begin{equation}
\label{eq:3}
I(R) = I_e\exp\{-b_n[(\frac{R}{R_e})^{1/n}-1]\},
\end{equation}
where $b_n$ satisfies:
\begin{equation}
\label{eq:4}
\gamma(2n; b_n)=\frac{1}{2}\Gamma(2n).
\end{equation}
where $R_e$ is the half-light radius, $I_e$ is the surface brightness at $R_e$, and $n$ is the S\'ersic index which reflects the steepness of the luminosity profile.\par

\begin{figure}
\epsscale{1.15}
\plotone{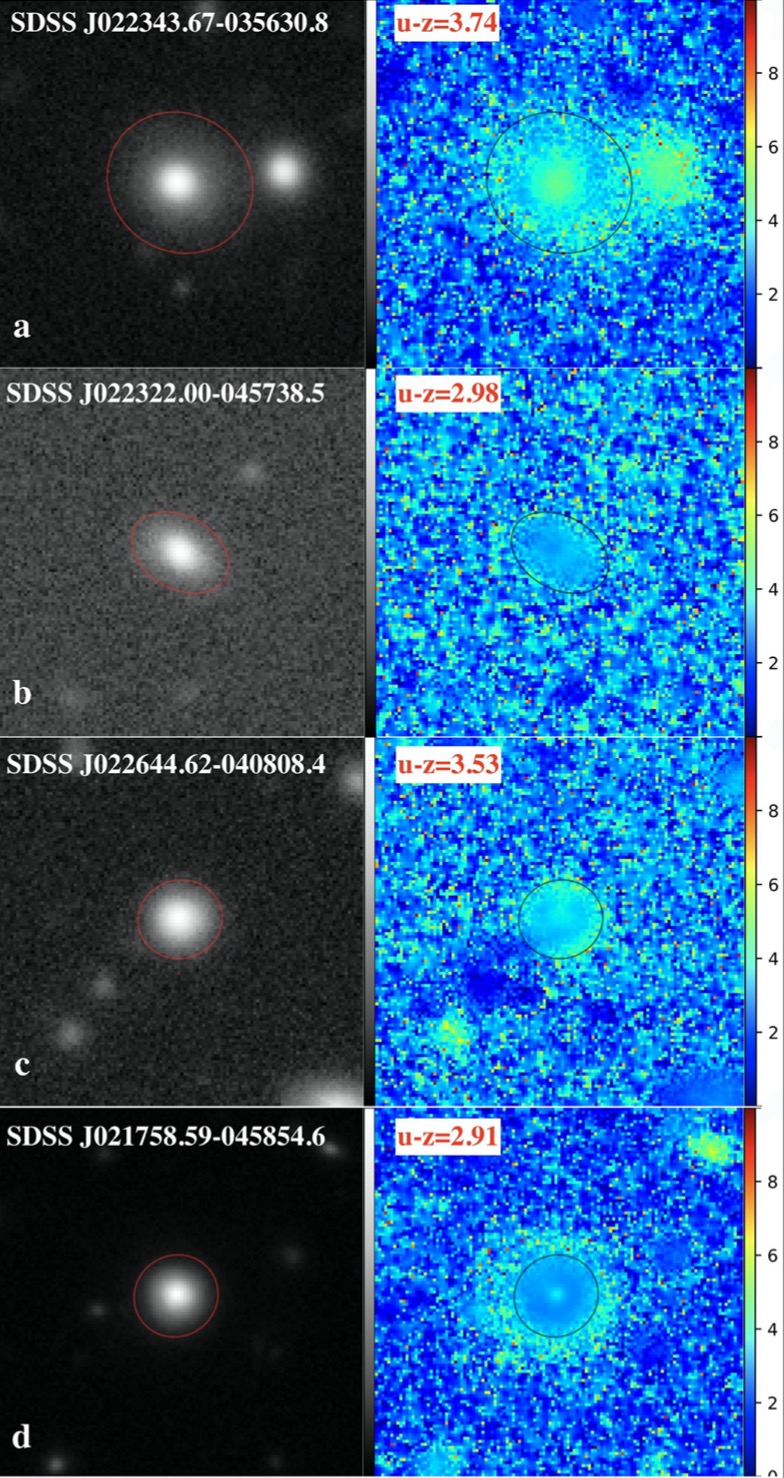}
\caption{The $z$ band images (left panels) and $u-z$ color maps (right panels) of four symmetric SF galaxies (Category III). Each panel covers a $20\arcsec\times20\arcsec$ area centered on each galaxy. Red (left panels) and black (right panels) ellipses indicate $z$ band Petrosian radii of the galaxies. \label{fig:f4}}
\end{figure}

We use \textsc{galfit}~\citep{2002AJ....124..266P} to derive $n$. At first, we mask other sources and build a mask map for each galaxy using \textsc{Sextractor}. Then, in order to fit the intrinsic profiles, we build the point spread function (PSF) of each galaxy through \textsc{PSFEx}  \citep{bertin2011automated}. In the configuration file of \textsc{galfit}, we input the mask maps, the PSF files, and set the initial values extracted from \textsc{Sextractor}. Finally, running \textsc{galfit} on the $z$ band luminosity maps of the galaxies, we get the S\'ersic indices for the four galaxies in Category III.\par

Generally, disk galaxies have exponential profiles, namely $n=1$~\citep{2011ARA&A..49..301V}. Galaxies that have $n$ larger than 2 should possess relatively big bulges.

The last 4 rows in Table \ref{tab:t1} show the properties of the four galaxies in Category III and Figure \ref{fig:f4} shows their images. The first one (Figure \ref{fig:f4}a) has S\'ersic index $n=3.04$, and has relatively lower SFR and redder color, indicating that it is an elliptical galaxy. The second one (Figure \ref{fig:f4}b) seems to be a disk with $n=1.61$. However, because of its relatively small $b/a$, its inner structure is hard to observe. It is hard to tell whether it has an asymmetric structure or not, so it may be a disk. The third one (Figure \ref{fig:f4}c) is a disk galaxy according to its low S\'ersic index $n=1.17$, which indicates a nearly exponential profile. Surprisingly, the fourth one (Figure \ref{fig:f4}d) has $n=2.32$, a very large SFR, and a very low $A_{abs}$. This galaxy looks more like an elliptical but with strong SF. And observed by eye, it truly has a more concentrated bulge than the third one.  A star forming disk with a concentrated bulge may be the product of a major wet merger, with the stellar component of the progenitors forming the bulge, and with the remaining cold gas bound to the descendant galaxy settling down and forming the SF disk. However, most of the blue ellipticals formed by wet mergers usually experience central starbursts, which lead to bluer bulges than their surrounding disks \citep{2009ApJS..182..216K}. The blue elliptical we find here has a bluer outer part than its bulge, which may imply that no strong starburst has happened at the bulge. The total $EW$ for emission lines of this galaxy is given in Table \ref{tab:t1} (the last row), which is measured within 3 arcsec diameter aperture in the galactic center. The small emission line $EW$ (6.5\AA) supports our analysis that there is no strong star formation at the bulge. 

We should caution that up to now we determined disk or elliptical galaxies only based on photometric images. We will use the Illustris simulation to study their formation characters and to determine whether they are true disks. \par

\section{Properties of Simulated Galaxies}
\label{sec:simulation}

By analyzing HSC+CLAUDS data in the previous section, we have found some correlation between $A_{abs}$ and the $u-z$ color, which indicates a possible connection between galaxy mergers and their SF activities in massive galaxies. In this section, we will analyze galaxies in the Illustris simulation in the same way as in observation, and use their merger histories to study the connection between mergers and SF.

\begin{figure*}
\epsscale{1.1}
\plotone{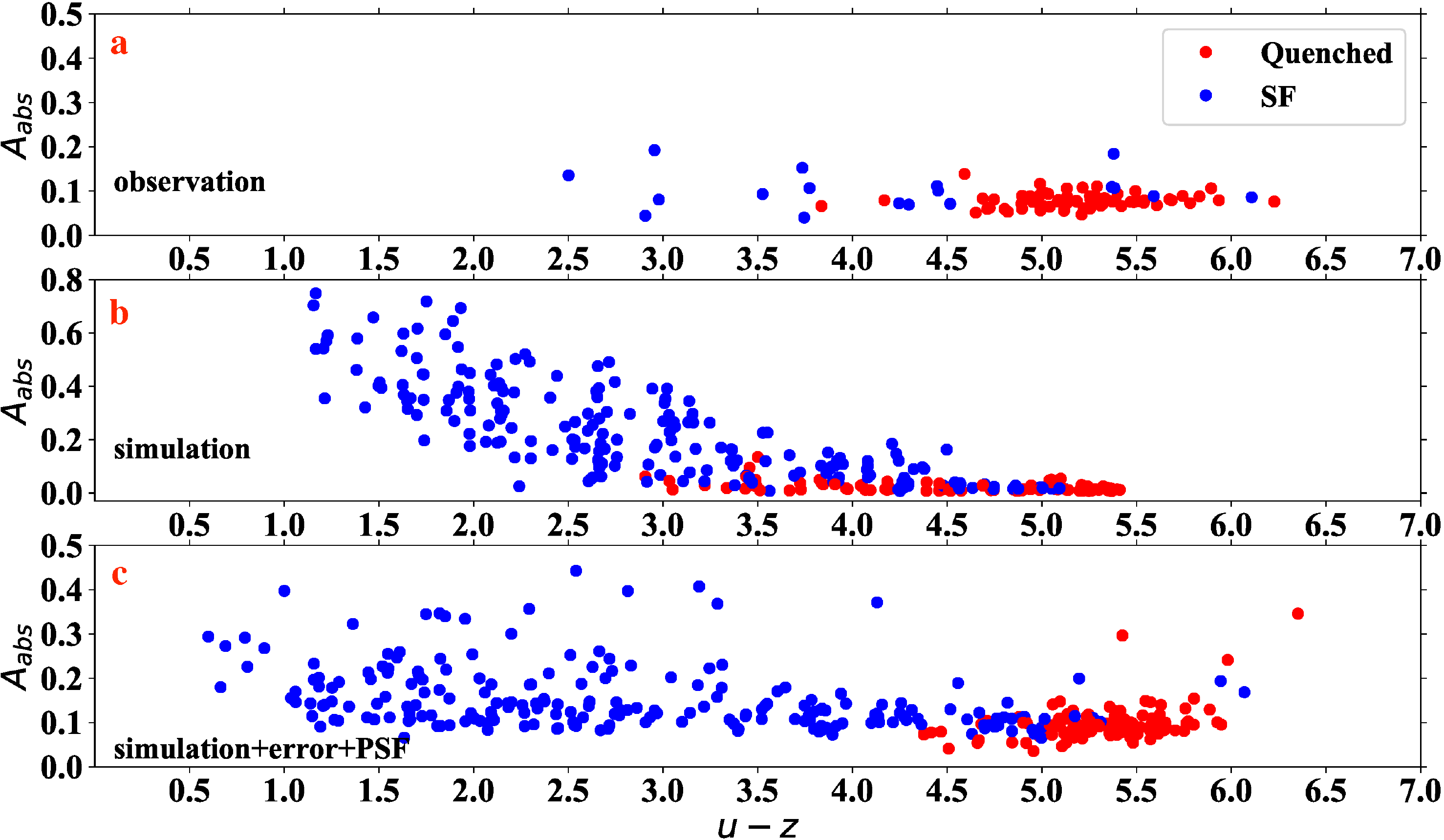}
\caption{Distributions of $A_{abs}$ versus $u-z$ for both the observation and the simulation. \textit{Panel (a):} Calculated using the observation images. \textit{Panel (b):} Calculated using the Illustris redshifted images without including observational shot noise and seeing effects. \textit{Panel (c):} Calculated using Illustris redshifted images after the observational shot noise and seeing effects have been mimicked . \label{fig:f5}}
\end{figure*}

\subsection{Asymmetry analysis} 
\label{subsec: asymmetry}

With Equation \ref{eq:1}, galaxies in Illustris can also be divided into SF galaxies and quenched galaxies. In this way, 41 out of the total 81 massive galaxies ($51\%$) in the Illustris are SF galaxies, which gives a much higher fraction of SF galaxies than the observational sample ($22\%$). The reason might be that the Illustris simulation has adopted an inefficient radio-mode AGN feedback that fails to quench enough massive galaxies as in the reality \citep{2014MNRAS.444.1518V}. Hence, we expect that there might be more symmetric and blue galaxies in the Illustris sample than in observation. \par

We use the redshifted images at $z=0.5$ (snapshot 103), which are provided by the Illustris team \citep{2015MNRAS.447.2753T}, to do the same analysis as in observation. The redshifted images were constructed by placing the galaxy 50 Mpc away from the observer. Hence, to mock the observational images, we perform dimming on the images and move the galaxies further to $z=0.5$. Then we unify the units and scale simulated images to the
CLAUDS/HSC pixel sizes using \textsc{Swarp}~\citep{bertin2010swarp}. Next, the Poisson noise is calculated from the noise images given by CLAUDS and HSC, and then is added to the simulation $u$ and $z$ band images. Finally, using \textsc{PSFEx}~\citep{bertin2011automated}, we build PSFs for 83 massive galaxies (including SF and quenched galaxies) in our observational sample. For each mock galaxy in the Illustris sample, we randomly select one of the 83 PSFs and make the convolution. We then get the similar $u-z$ maps to the observed ones, and calculate their $A_{abs}$ in the same way as in the observation. Since Illustris provides images in four lines of sight for each galaxy, we regard them as four independent galaxies in the analysis of galaxy asymmetry. \par

Figure \ref{fig:f5} shows the relation between galaxy $u-z$ color and $A_{abs}$ for both the observational and the simulation samples. We find that Illustris has many more blue galaxies than the observational sample. In detail, Illustris has many galaxies with $u-z$ bluer than 2.5 while none were found in the observational sample. It is likely caused by the inefficient radio-mode AGN feedback in the Illustris simulation model as mentioned above. If we truncate the Illustris sample at $u-z=2.5$, the trends in observation and simulation are very similar. There is also some correlation between $u-z$ color and $A_{abs}$ in Illustris, which is consistent with what we found for the observational sample. Furthermore, SF galaxies in Illustris also have higher asymmetry than quenched ones at given color. Figure \ref{fig:f5}b shows the asymmetry distribution of the original Illustris images, which reflects the true properties of the galaxies. In this panel, the difference between the SF galaxies and the quenched galaxies becomes much more evident. Nearly all quenched galaxies are symmetric and all SF galaxies have relatively high $A_{abs}$. The intrinsic correlation between the $u-z$ color and $A_{abs}$ is even stronger. Since the mock images of the truncated Illustris sample ($u-z>2.5$) have a similar trend as in the observation, we expect that the true underlying SF-asymmetry correlation for the observations is similar to the distribution shown in the Figure \ref{fig:f5}b. Indeed, the observational effects, both the seeing and the shot noise, have significantly degraded the correlation, as the seeing effect smears the asymmetry of blue galaxies and the shot noise in $u$ band boosts the asymmetry in quenched galaxies. Comparing the Figure \ref{fig:f5}b and \ref{fig:f5}c, we can see that nearly all quenched galaxies have much larger $A_{abs}$ after the shot noise is included, and some of them have been boosted a large amount by the shot noise. This also implies that the $A_{abs}$ measurements of SF morphology is only meaningful for SF galaxies in practice. \par

\begin{figure*}
\epsscale{0.55}
\plotone{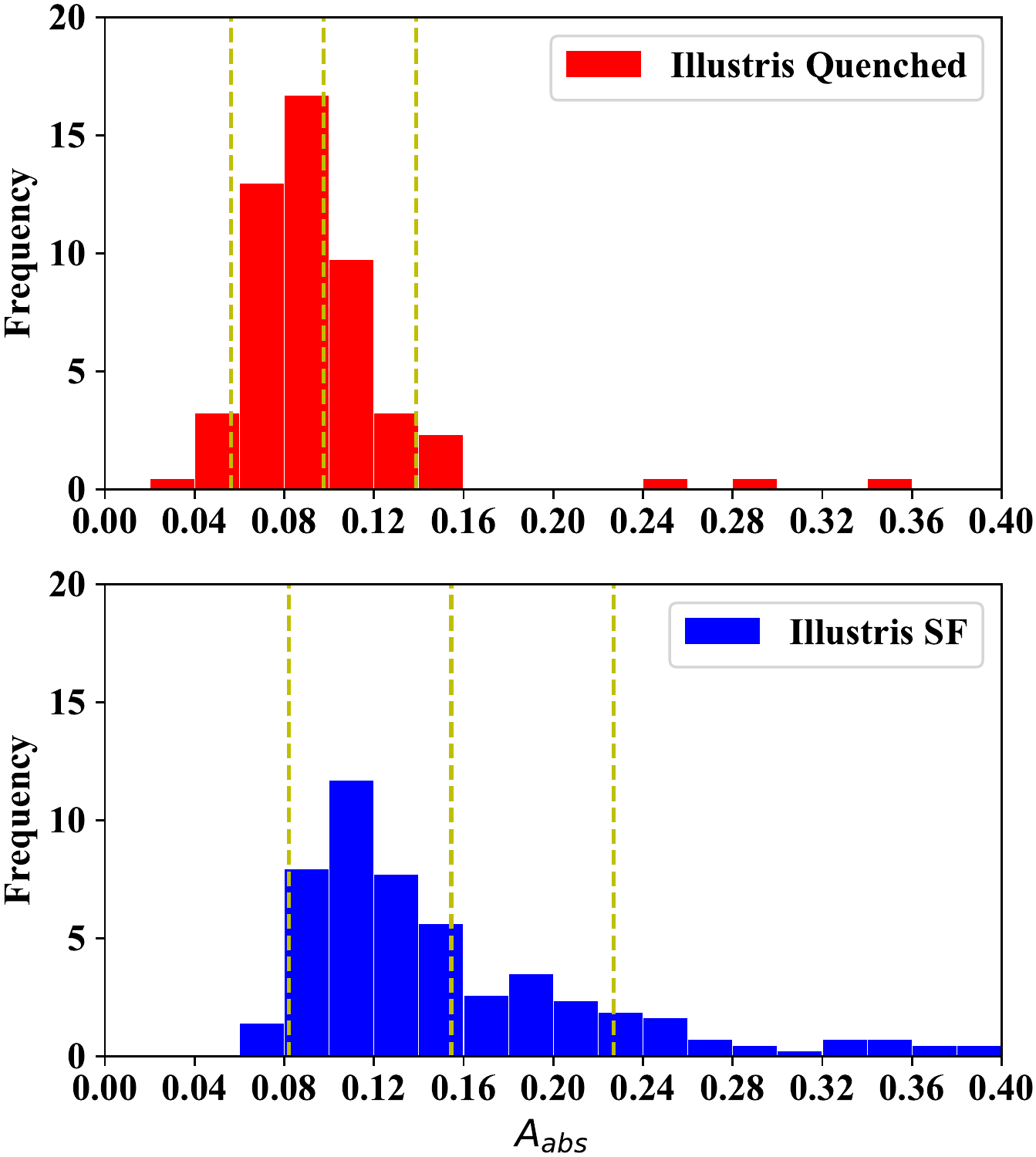}
\epsscale{0.55}
\plotone{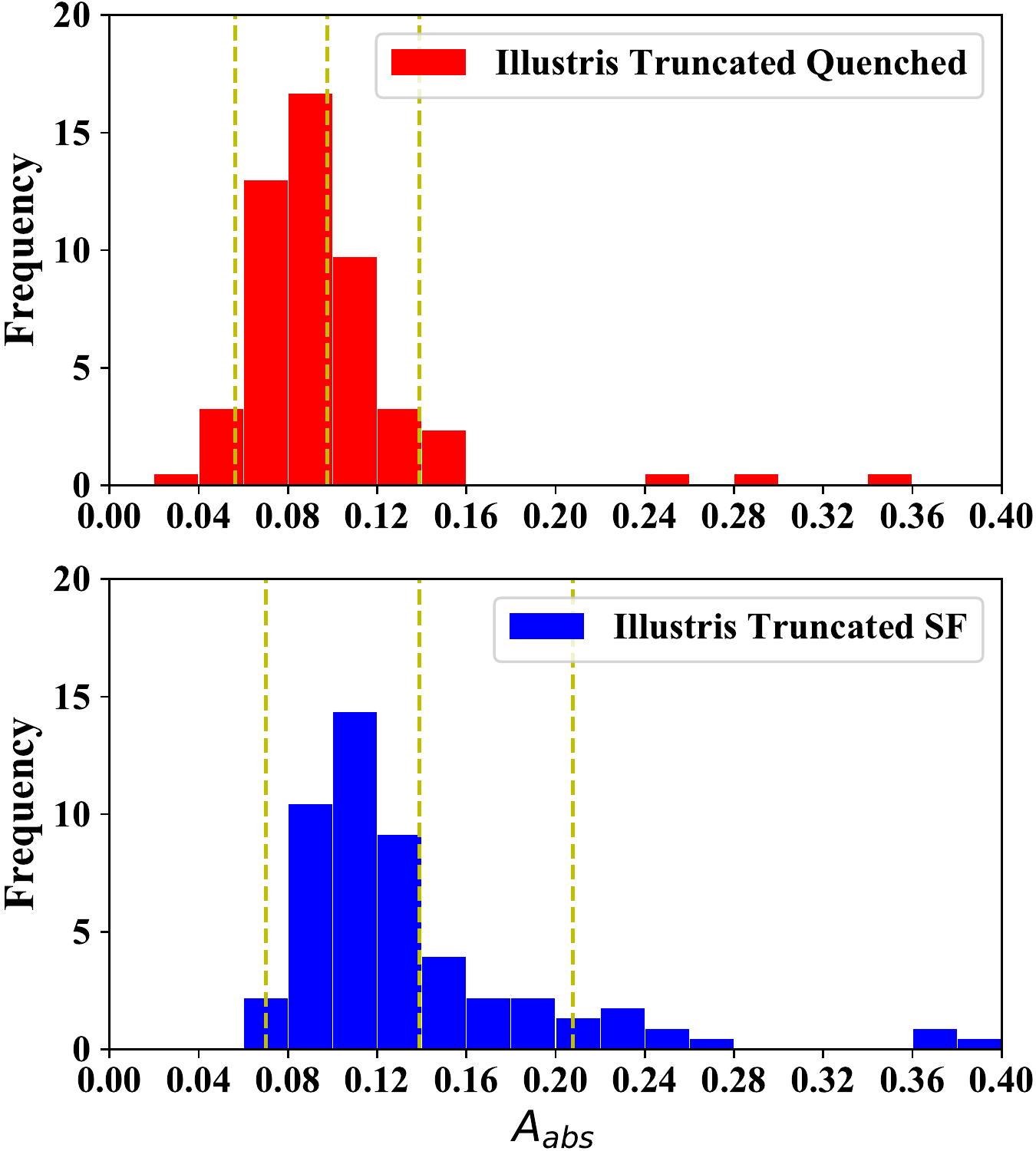}\\
\caption{Distributions of $A_{abs}$ for quenched (upper panels) and  SF (lower panels) galaxies with the mean values and 1$\sigma$ deviations shown as vertical dashed lines in each panel. As in Figure \ref{fig:f3}, we calculate KS-test \textit{p}-values for the $A_{abs}$ distributions between SF and quenched galaxies. \textit{Left panels:} Normalized $A_{abs}$ distributions for the Illustris sample with the KS-test \textit{p}-value $3.712\times10^{-17}$. \textit{Right panels:} Normalized $A_{abs}$ distributions for the Illustris sample truncated at $u-z=2.5$ with the KS-test \textit{p}-value $1.659\times10^{-10}$. \label{fig:f6}}
\end{figure*}

We show the $A_{abs}$ distributions of quenched (upper panels) and SF (lower panels) galaxies in Figure \ref{fig:f6} for the whole Illustris sample (left panels), and the Illustris truncated sample (right panels). As we did for the observational sample (see Figure \ref{fig:f3}), we apply K-S test to the $A_{abs}$ distributions for quenched and SF galaxies in Illustris sample. The K-S $p$-value is $3.712\times10^{-17}$ for the whole Illustris sample and $1.659\times10^{-10}$ for the Illustris truncated sample ($u-z>2.5$), which means that the $A_{abs}$ distribution of SF galaxies are significantly different from that of quenched galaxies. As we can see in Figure \ref{fig:f6}, SF galaxies tend to have higher $A_{abs}$ (i.e., more asymmetric), which is consistent with what we found in the observational galaxy sample. The result implies that mergers play important roles in the SF activities of the massive galaxies.

\subsection{Merger history} 
\label{subsec:merge history}

It is convenient to trace the merger histories of galaxies in Illustris using the \textsc{sublink} merger trees. Thus, we can demonstrate the relation between SF and galaxy mergers in massive galaxies explicitly through the merger trees. Each \textsc{sublink} merger tree is identified with a `main progenitor branch' where the algorithm follows the subhalo progenitor branch with the `most massive history'~\citep{2007MNRAS.375....2D}. Galaxies on the main progenitor branch are defined as the main galaxies, and the galaxies which merge onto the main progenitor branch are defined as the infall galaxies. The time of each merger event is considered to be the time of the snapshot in which the infalling galaxy was last independently identified.

\begin{figure*}
\gridline{\fig{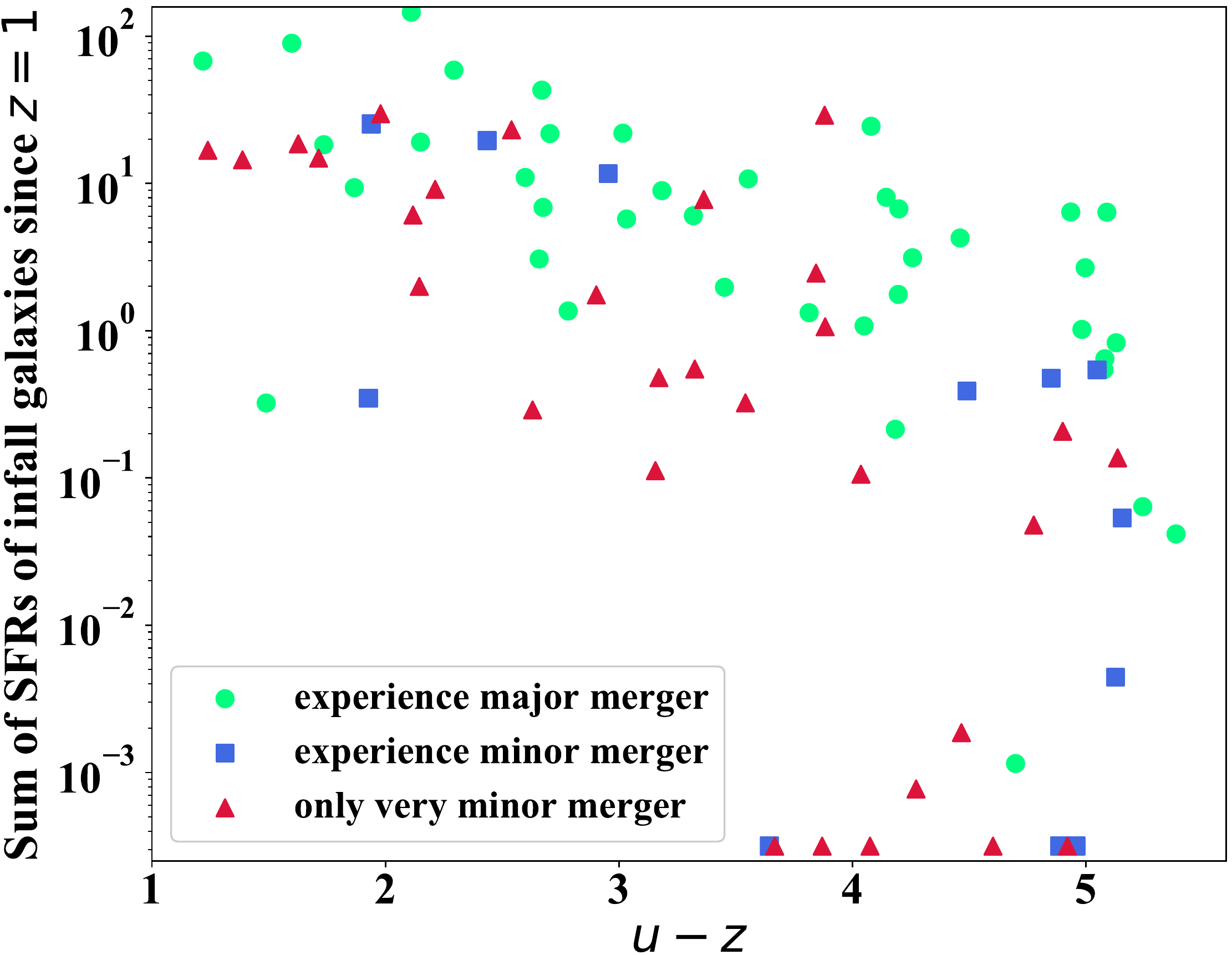}{0.48\textwidth}{(a)}\fig{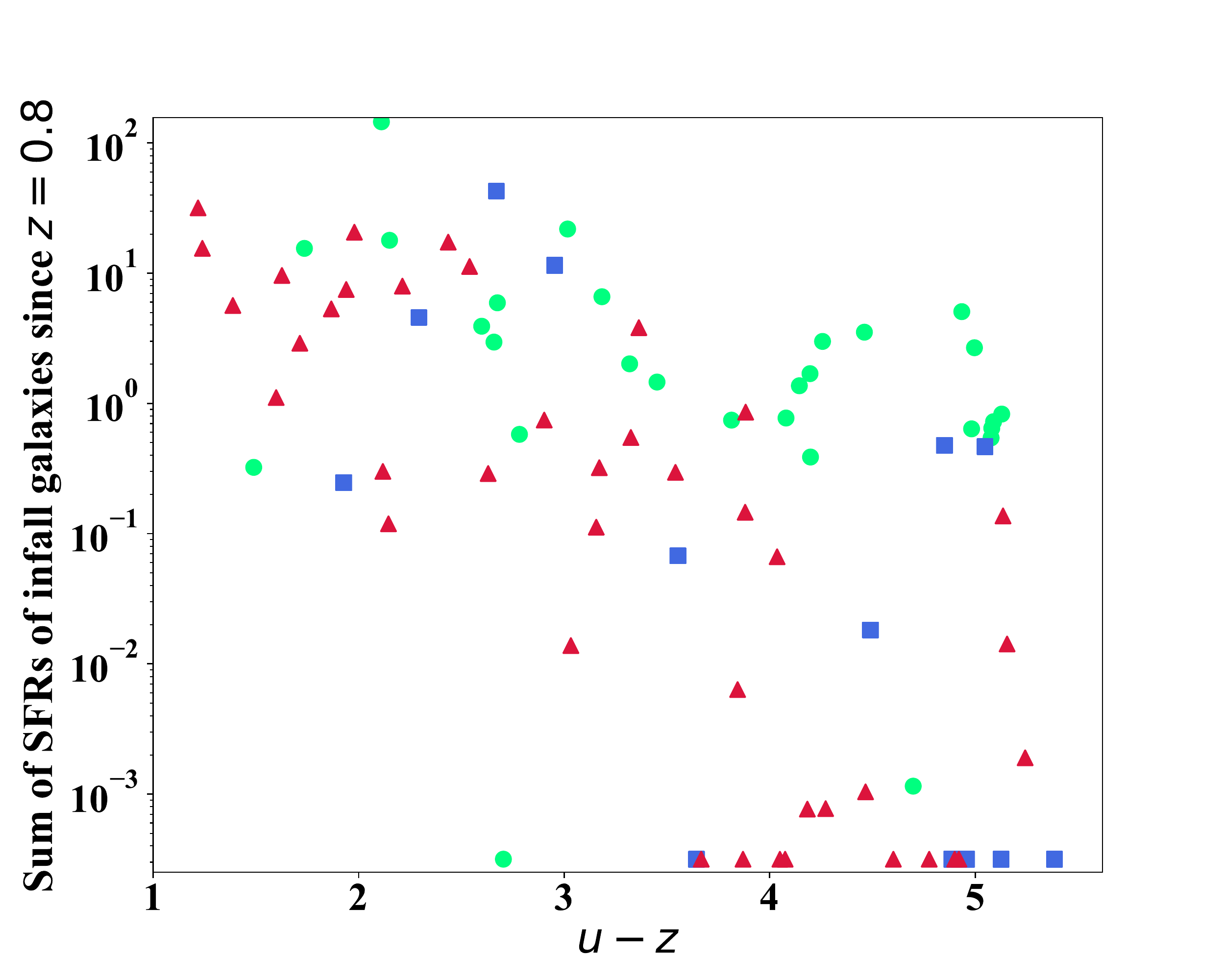}{0.48\textwidth}{(b)}}
\caption{We plot the sum of SFRs of infall galaxies at their merger time as a function of the $u-z$  colors (mean value of the four lines of sight) of the main galaxies at $z=0.5$. SFRs are in logarithmic scale. Galaxies that have experienced major mergers are plotted in green filled circles, while those that haven't experienced major but minor mergers are plotted in blue filled squares. The galaxies that only experienced very minor mergers are plotted in red filled triangles. \textit{Panel (a):} Sum of the SFRs of the galaxies that merged since $z=1$. \textit{Panel (b):} the same as \textit{Panel (a)}, but since $z = 0.8$. \label{fig:f7}}
\end{figure*}

Figure \ref{fig:f7} shows the relation between the $u-z$ colors of the 81 main galaxies at redshift $z_r=0.5$ and the sum of SFRs of the infall galaxies at their merger time in the intervals: $0.5<z_r<0.8$ and $0.5<z_r<1.0$. It presents how much {\it ex situ} SF of recently merged galaxies has contributed to the colors of the main galaxies. To be more rigorous, we should use the sum of the time-weighted SFRs~\citep{2014ARA&A..52..415M}. However, since the results are already clear and obtaining the SFR evolution function requires a lot of assumptions, we just perform the simple sum of SFRs in the two-time intervals. The different colors and symbols in the plots represent three kinds of mergers according to different merger mass ratios $\mu$. The definition is: \\
\indent $\bullet$ major merger: $\mu > 1/4$ ;\\
\indent $\bullet$ minor merger: $1/10 < \mu \leq 1/4$;\\
\indent $\bullet$ very minor merger: $\mu \leq 1/10$.\\
Specifically, $\mu$ is defined as the stellar mass ratio of the infalling galaxy to the main galaxy when the infalling galaxy historically reached its maximum stellar mass~\citep{2015MNRAS.449...49R}. We use the SFRs of the infall galaxies at the merger snapshots, i.e. immediately before the merger process.\par

In Figure \ref{fig:f7}a and \ref{fig:f7}b, it is easy to see that the sum of {\it{ex situ}} SFRs has a clear correlation with the $u-z$ colors of the main galaxies, which proves that recent mergers do contribute significantly to the SF of the main galaxies at a later time (i.e. at $z=0.5$). Interestingly, we can see from Figure \ref{fig:f7}a and \ref{fig:f7}b that the different sets of merger types follow similar trends of the sum of SFRs vs the $u-z$ color, which indicates that the color of the massive galaxies is only dependent on the sum of the SFRs of infall galaxies, independent of the merger types. \par

In conclusion, using the merger trees, we find that the $u-z$ colors of the main galaxies in our Illustris sample show a significant dependence on the sum of SFRs of recent infall galaxies at their merger time. This could be interpreted as the result that the newly formed stars of the infall galaxies contributing to the colors and SFRs of the main galaxies.\par

\subsection{Disk galaxies } 
\label{subsec:disk galaxies}

\begin{deluxetable}{ccccccc}
\tabletypesize{\scriptsize}
\tablenum{2}
\tablecaption{Properties of the seven simulated galaxies with $n<2.0$.\label{tab:t2}}
\tablewidth{0pt}
\tablehead{
\colhead{Subhalo ID}&\colhead{M$_\ast$}&\colhead{$D/T$}&\colhead{SFR}&\colhead{$A_{abs}$}&\colhead{$n$}&\colhead{$u-z$}\\
\colhead{}&\colhead({M$_{\odot}$)}&\colhead{}&\colhead{(M$_{\odot}\,$yr$^{-1}$)}&\colhead{}&\colhead{}&\colhead{(mag)}
}
\startdata
109879&$10^{11.45}$&0.065&$10^{1.3}$&0.10&$1.59$&2.58\\
118801&$10^{11.64}$&0.131&$10^{0.9}$&0.12&$1.10$&3.51\\
155335&$10^{11.45}$&0.235&$10^{1.4}$&0.14&$1.10$&2.20\\
107679&$10^{11.49}$&0.299&$10^{1.7}$&0.14&$0.96$&1.22\\
183862&$10^{11.31}$&0.429&$10^{1.3}$&0.10&$1.10$&1.68\\
159685&$10^{11.53}$&0.491&$10^{1.3}$&0.13&$1.93$&2.38\\
143882&$10^{11.31}$&0.538&$10^{1.3}$&0.20&$1.20$&1.57\\
\enddata

\end{deluxetable}


In the observational sample, we have found two candidates of blue disk galaxies with the S\'ersic index $n<2$, and another candidate for a star forming spheroid of $n=2.3$. As Illustris has inefficient radio-mode AGN feedback, we expect more blue disks/spheroids in the simulated sample.\par

Since the Illustris provides kinematic information for every star particle, the classification of the disk and elliptical galaxies can be achieved in both kinematic and photometric means. Many previous studies \citep[e.g.,][]{2012ApJS..198....2K,2016ASSL..418..431K,2019MNRAS.487.5416T} have shown that these two methods sometimes lead to different classification results. Since we do not have kinematic data for galaxies in the observational sample and have classified them based on their luminosity profiles, we will use the S\'ersic index to classify disk and elliptical galaxies in the Illustris sample for self-consistency. But we will cross-check the classified results with the kinematic data, and study their relations to the merger histories as well.\par

\begin{figure*}
\gridline{\fig{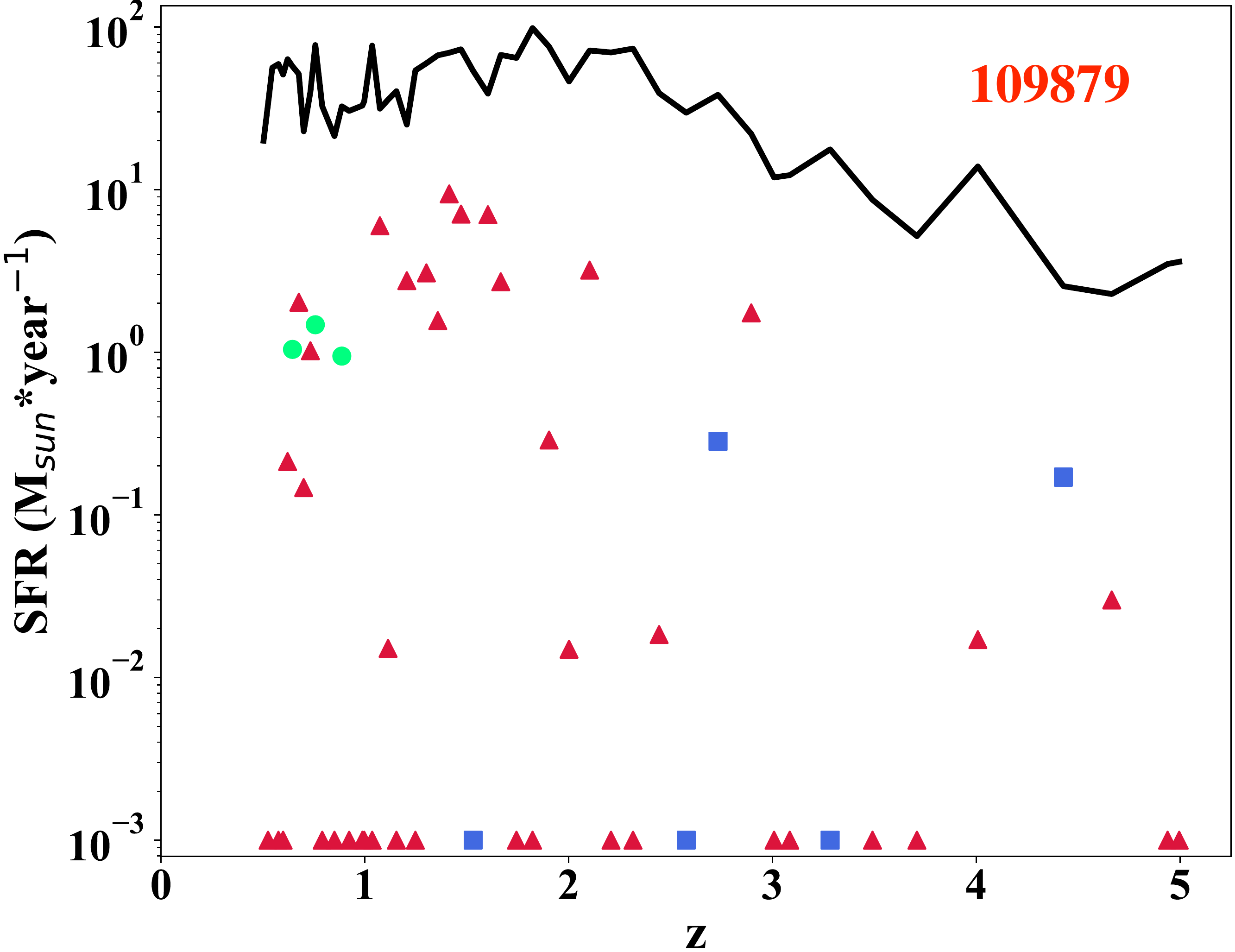}{0.33\textwidth}{(a)}\fig{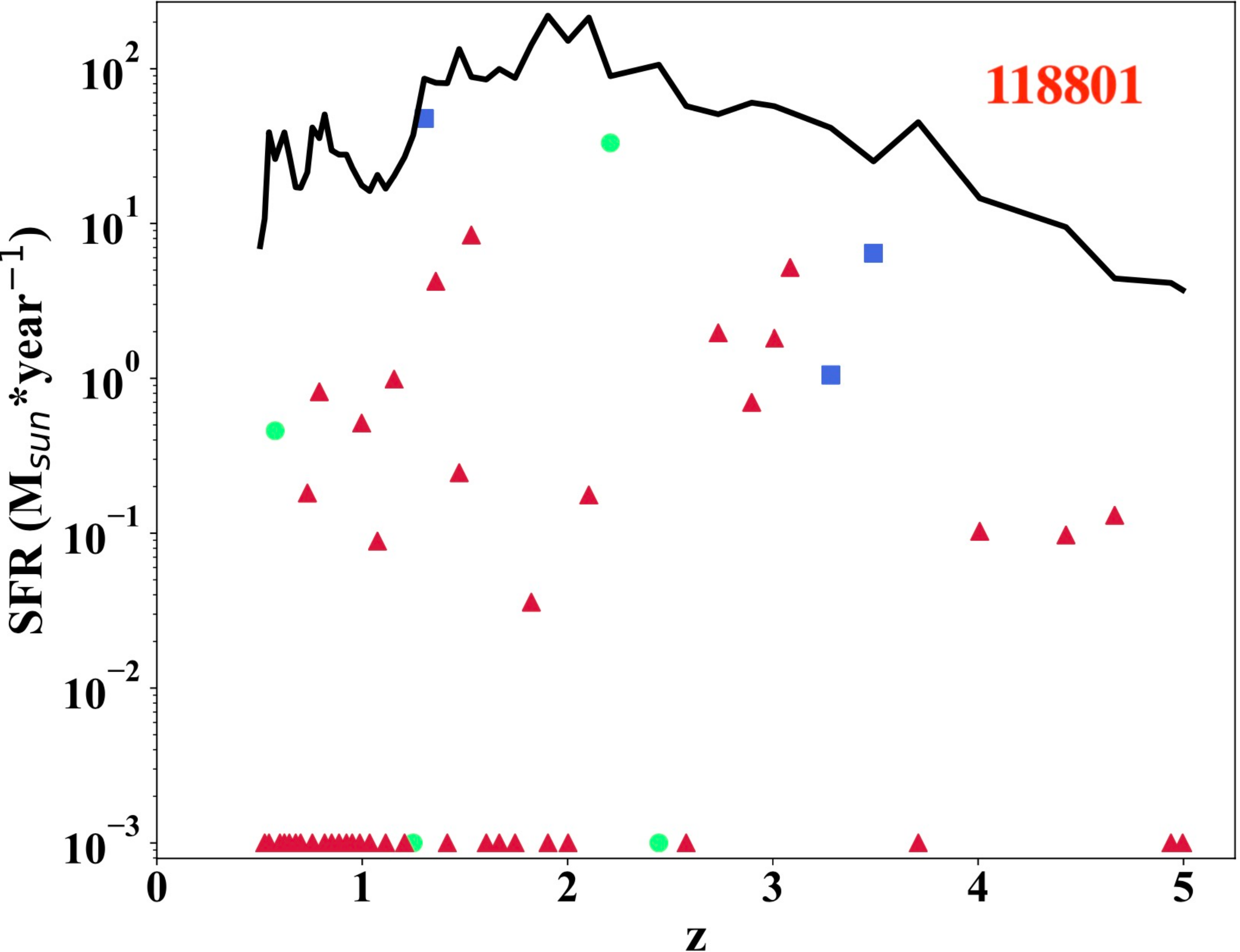}{0.33\textwidth}{(b)}\fig{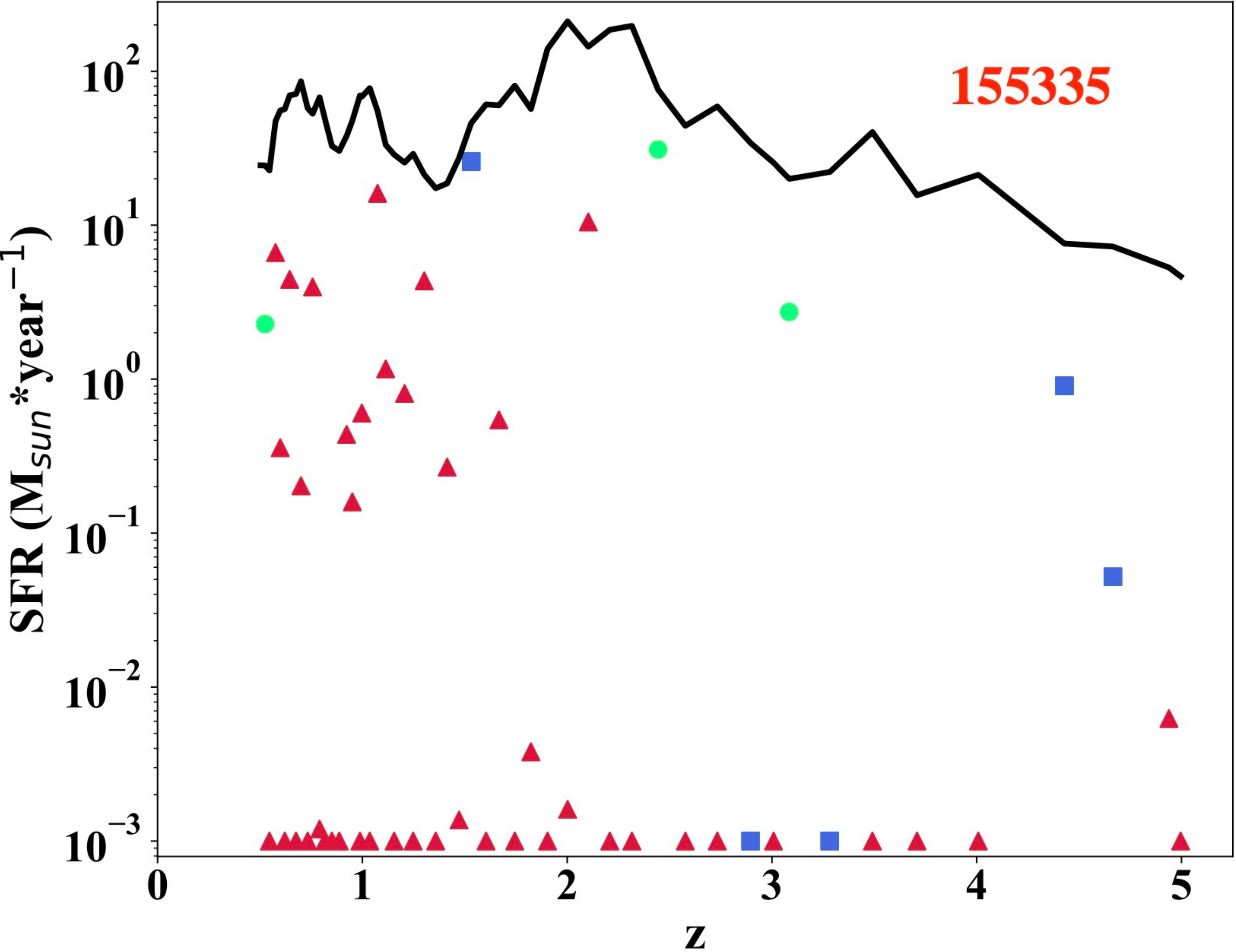}{0.33\textwidth}{(c)}}
\gridline{\fig{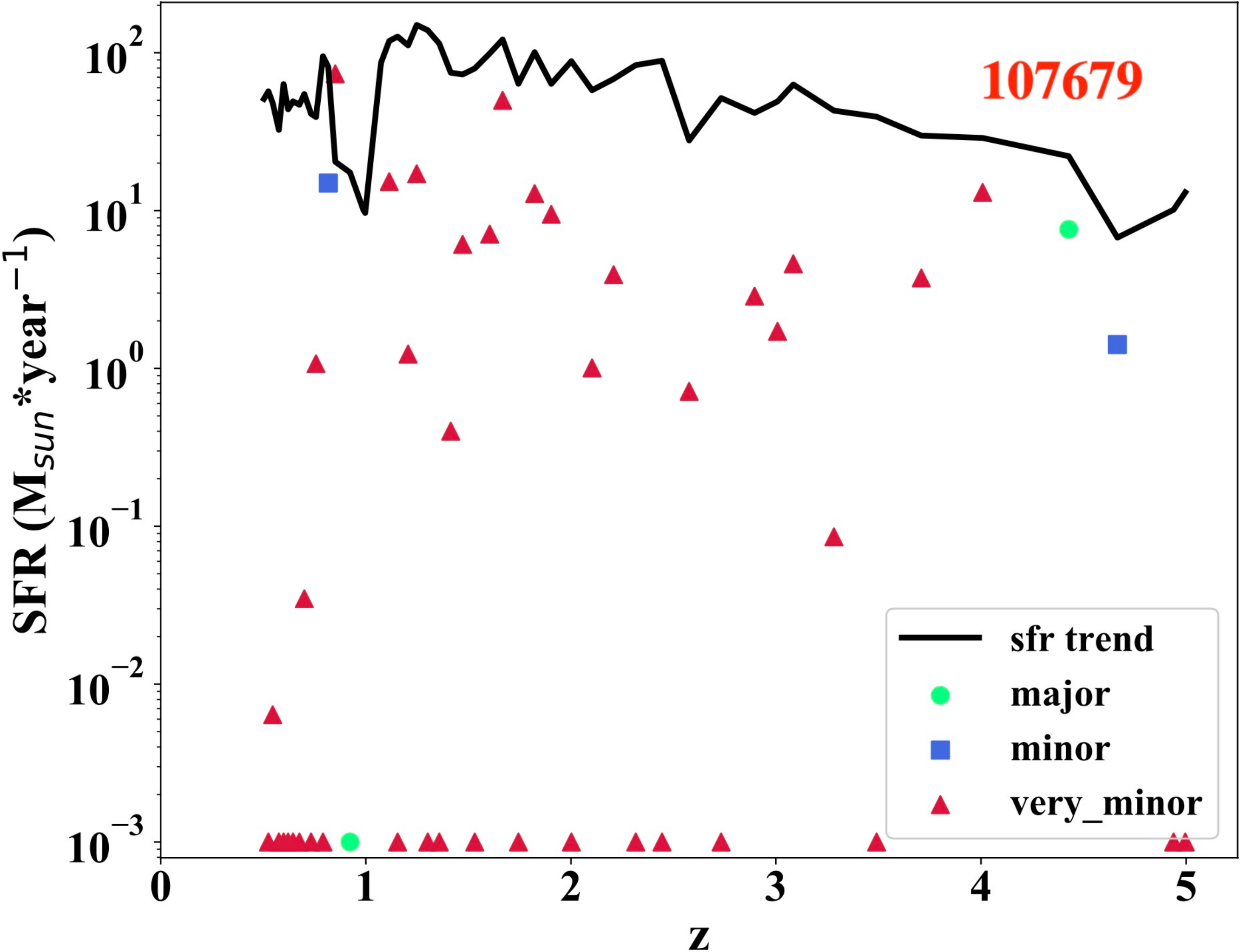}{0.33\textwidth}{(d)}\fig{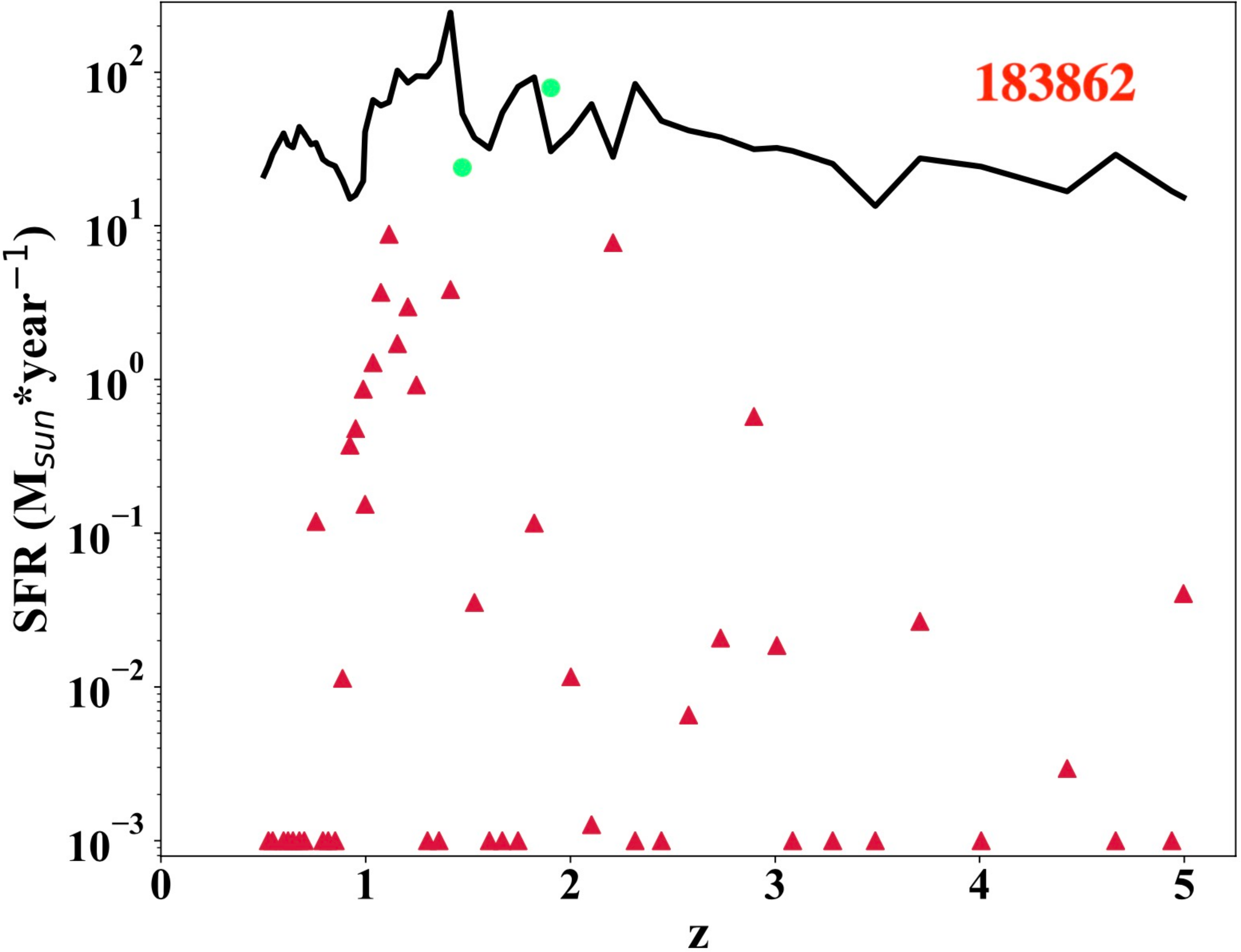}{0.33\textwidth}{(e)}\fig{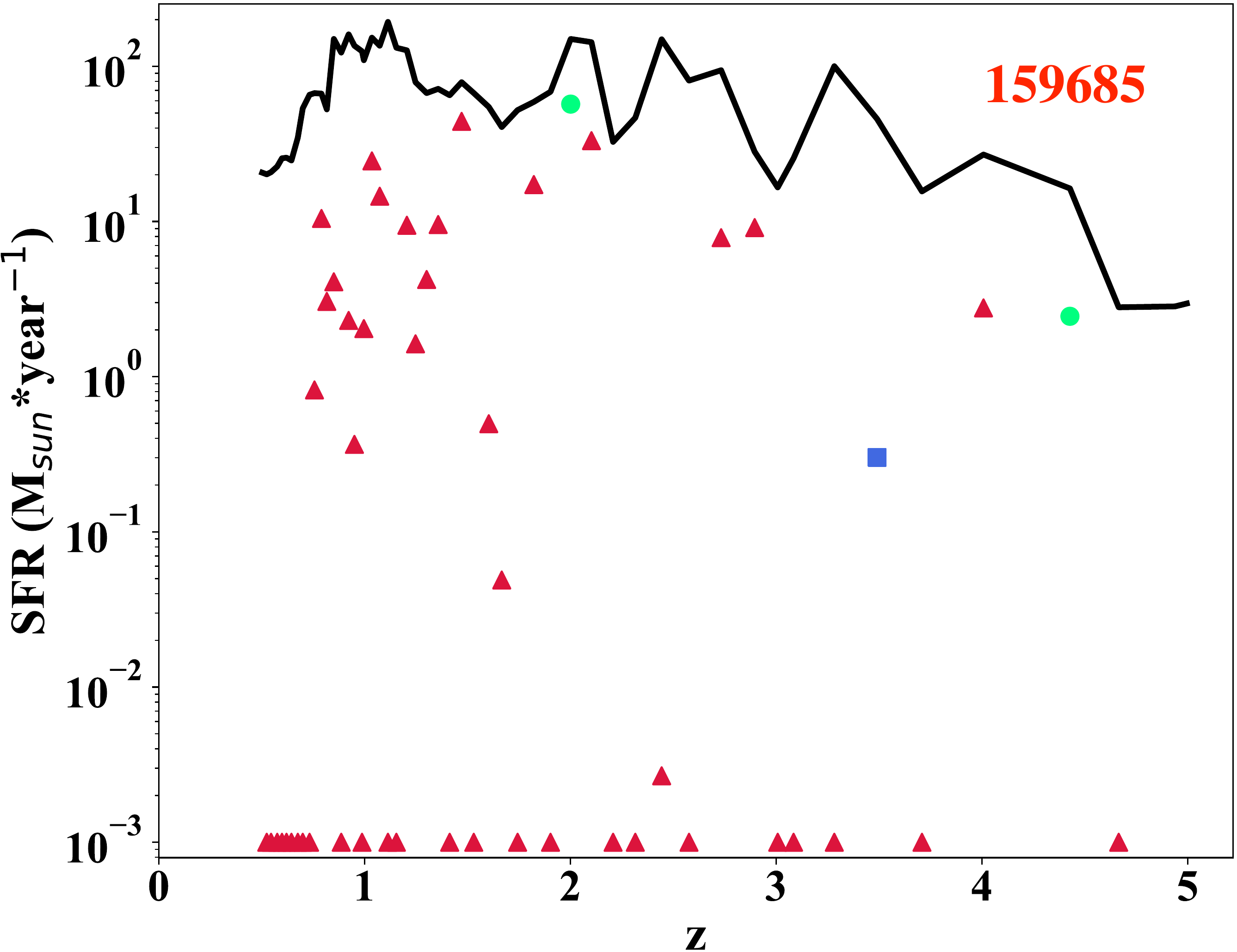}{0.33\textwidth}{(f)}}
\gridline{\fig{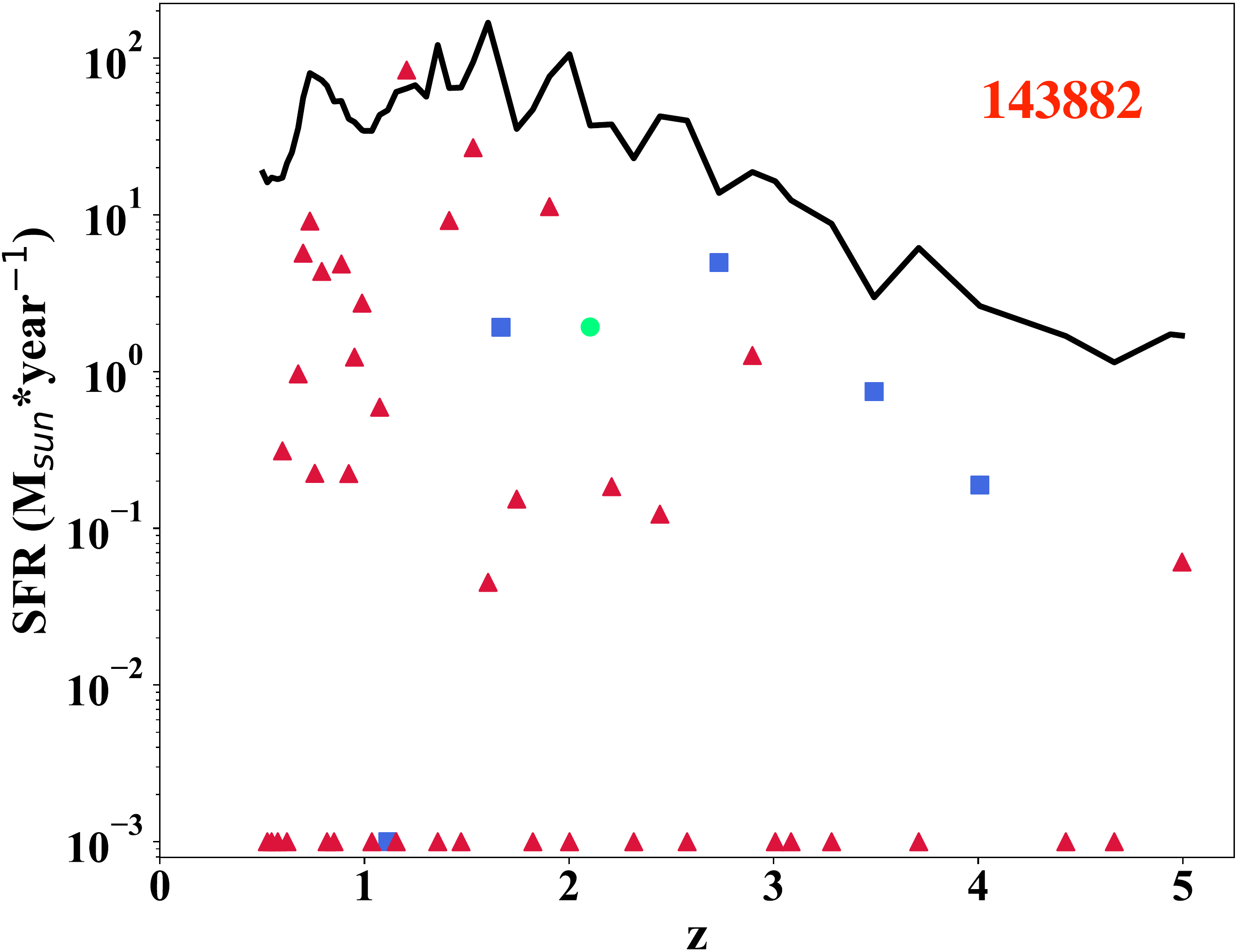}{0.33\textwidth}{(g)}}
\caption{Merger histories of the seven disk candidate galaxies with S\'ersic index $n<2$ in the Illustris. The black lines are the SFR of the main galaxies, and the colored dots show the SFR of the infall galaxies at their merger time. Different colors and symbols indicate different types of mergers, in the same way as in Figure \ref{fig:f6}. The subhalo ID of each galaxy is at the upper right corner of each panel.  \label{fig:f8}}
\end{figure*}

The Illustris provides the S\'ersic index in three directions for each galaxy in its data release~\citep{2017MNRAS.469.1824X}. Here we use the maximum S\'ersic index of the three lines of sight. There are a totally 12 SF galaxies that have $n<2.0$. As for the observational sample, we visually check $z$ band images of these 12 SF galaxies and find that five of them show clear merger signatures. Excluding these five galaxies,  we get seven galaxies as disk candidates finally. Then we check their kinematic properties using the disk-to-total ratio $D/T$ which is defined as the ratio of the number of disk stellar particles to the number of all stellar particles in a galaxy \citep{2015ApJ...804L..40G}. Generally, galaxies of $D/T>0.3$ are defined as disks in many studies~\citep{2014MNRAS.444.1518V,2015ApJ...804L..40G,2019MNRAS.487.5416T}.  Under this definition, of the above seven galaxies, four are ellipticals and three are disks. Table \ref{tab:t2} gives the properties of these seven galaxies sorted by $D/T$. $A_{abs}$ and $u-z$ here are the mean values of these two quantities along the four lines of sight. In comparison with observation, disk galaxies in Illustris have larger asymmetries and lower $u-z$ colors. It might be caused by the spiral arms and the overall higher SFR. Though in observation and simulation the color distributions of the galaxies are different, we think the mechanism to maintain a disk is similar. Major mergers can quickly destroy the disk structures and transform disk galaxies into elliptical ones. Thus, the disk galaxies should not have been through major mergers, and thus have maintained their disks. To justify this, we study the merger histories of the seven candidates of blue disk galaxies in the Illustris.\par

We show the individual merger histories of the seven galaxies in Figure \ref{fig:f8}. The first four galaxies all have experienced major mergers since $z=1$, which leads to small $D/T<0.3$. These major mergers can be hardly detected in a photometric way at $z=0.5$, but the infallen galaxies have not fully settled, which might lead to a small $n$ of the S\'ersic profile fitting. This is also the reason for some mismatch cases in the photometric and kinematic classifications. The other three galaxies with $D/T>0.3$ have not experienced any major mergers since $z\approx 1.5$, so they can maintain their disk structures. In these circumstances, the infallen galaxies have fully settled down in such a long time and the S\'ersic fitting reveals the true properties. Our finding on the relation between the $D/T$ and the merger histories reconfirm the expectation that the kinematic data are more reliable for identifying disk galaxies.\par

In conclusion, after removing galaxies that have clear merger signals in the $z$ band images, we get seven galaxies with $n<2$ in Illustris sample. However, only 3 of them are true disks according to their kinematic properties $D/T>0.3$, all of which have not been through major mergers since $z \approx 1.5$.  Though the $u-z$ distributions are different in the observation and the simulation, we believe they share the common mechanism for maintaining disks. From the simulation, we would caution that the disk candidates identified based on the S\'ersic fitting may not be true disks. If the fraction of true disks in the Illustris is typical, we would expect that about half of the blue disk candidates found in the observation based on the S\'ersic fitting are true disks. It would be interesting to carry out the Integral Field Unit (IFU) observations for the candidate disks found in the last section. Finally, we would like to point out that we have not found any galaxies of $ n>2$ in the simulation that is forming stars as strong as while having color as blue as the observed spheroid (the last galaxy in Table \ref{tab:t1}). \par

\section{Conclusions}
\label{sec:conclusion}

In this paper, we have analyzed the SF activity regions and morphology of a sample of galaxies with stellar mass $M_*>10^{11.3} {\rm M_{\odot}}$ at $z \sim 0.5$ using the HSC+CLAUDS survey data and the Illustris simulation data. Our main results are as follows:

\begin{itemize}
\item Using Equation~\ref{eq:1}, we divide galaxies into SF and Quenched galaxies. In the observational sample, 18 out of 83 galaxies are SF galaxies ($\sim 20\%$). However, 41 out of 81 galaxies are SF ones in the Illustris simulation. The difference in the fraction may be attributed to the overall high SFR in the Illustris caused by the inefficient radio-mode AGN feedback in the simulation model that fails to quench many massive galaxies.

\item From our visual inspection, we have identified 14 SF galaxies in the observed SF sample that show clear asymmetric structures in their $u-z$ color maps, and we are able to identify recent or/and undergoing mergers in most of their $z$ band maps (the stellar distribution). Of the rest four galaxies that show symmetric star formation and stellar distribution, three are in vigorous star formation ({\it in situ}).

\item Using the quantitative asymmetry parameter $A_{abs}$, we have studied the relation between the $u-z$ color and $A_{abs}$ in both the observation and the simulation. Excluding the four symmetric galaxies in the observational sample, we find a correlation between the $u-z$ color and $A_{abs}$. With the simulation, we are able to qualitatively reproduce the correlation. By analyzing the merger histories of the galaxies, we find that the correlation is a consequence of recent mergers. Combined with our visual inspection results, we find about 80\% of the SF massive galaxies are connected to recent mergers. 

\item Based on the S\'ersic profile analysis, we find that two symmetric SF are good candidates for blue disks and the other is a candidate for a SF spheroid. By jointly analyzing the morphology, kinematics, and merger histories of galaxies in the simulation, we find that the disk candidates have about 50\% chance to be real disks. We have not found any analogs in the simulation to the massive SF spheroid we found.

\end{itemize}

In summary, we find that about ~20\% of the massive galaxies at $z\sim 0.5$ are SF galaxies. In about 85\% of the SF galaxies, the SF activities are mainly induced by mergers. The rest of the SF galaxies (i.e. about 15\% of SF galaxies or about 3\% of massive galaxies) do not show significant merger signatures. These symmetric SF galaxies are likely massive blue disks or blue spheroids that are being formed vigorously. The formation of such massive disk/spheroid galaxies has interesting implications for the current theory of galaxy formation, especially for the feedback model.  We plan to carry out IFU observations for these symmetric SF galaxies. Furthermore, the current study is still limited by the sample size, and we also plan to extend the study to a larger sample of massive galaxies.

\acknowledgements
The work is supported by NSFC (11533006, 11890691, 11673017, 11933003, 11833005). We are grateful to Xiaoyang Xia, Caina Hao, Yongzhong Qian, and Yanmei Chen for their helpful discussions. 

This research uses data obtained through the Telescope Access Program (TAP), which has been funded by the National Astronomical Observatories, Chinese Academy of Sciences, and the Special Fund for Astronomy from the Ministry of Finance. 
This paper uses data from the VIMOS Public Extragalactic Redshift Survey (VIPERS). VIPERS has been performed using the ESO Very Large Telescope, under the "Large Programme" 182.A-0886. The participating institutions and funding agencies are listed at http://vipers.inaf.it. The Hyper Suprime-Cam (HSC) collaboration includes the astronomical communities of Japan and Taiwan, and Princeton University. The HSC instrumentation and software were developed by the National Astronomical Observatory of Japan (NAOJ), the Kavli Institute for the Physics and Mathematics of the Universe (Kavli IPMU), the University of Tokyo, the High Energy Accelerator Research Organization (KEK), the Academia Sinica Institute for Astronomy and Astrophysics in Taiwan (ASIAA), and Princeton University. Funding was contributed by the FIRST program from Japanese Cabinet Office, the Ministry of Education, Culture, Sports, Science and Technology (MEXT), the Japan Society for the Promotion of Science (JSPS), Japan Science and Technology Agency (JST), the Toray Science Foundation, NAOJ, Kavli IPMU, KEK, ASIAA, and Princeton University. This publication has made use of data products from the Sloan Digital Sky Survey (SDSS). Funding for SDSS and SDSS-II has been provided by the Alfred P. Sloan Foundation, the Participating Institutions, the National Science Foundation, the U.S. Department of Energy, the National Aeronautics and Space Administration, the Japanese Monbukagakusho, the Max Planck Society, and the Higher Education Funding Council for England. Based on observations obtained with MegaPrime/MegaCam, a joint project of CFHT and CEA/IRFU, at the Canada-France-Hawaii Telescope (CFHT) which is operated by the National Research Council (NRC) of Canada, the Institut National des Science de l'Univers of the Centre National de la Recherche Scientifique (CNRS) of France, and the University of Hawaii. This work is based in part on data products produced at Terapix available at the Canadian Astronomy Data Centre as part of the Canada-France-Hawaii Telescope Legacy Survey, a collaborative project of NRC and CNRS.

\appendix

\begin{figure*}[h]
    \epsscale{1.15}
    \plotone{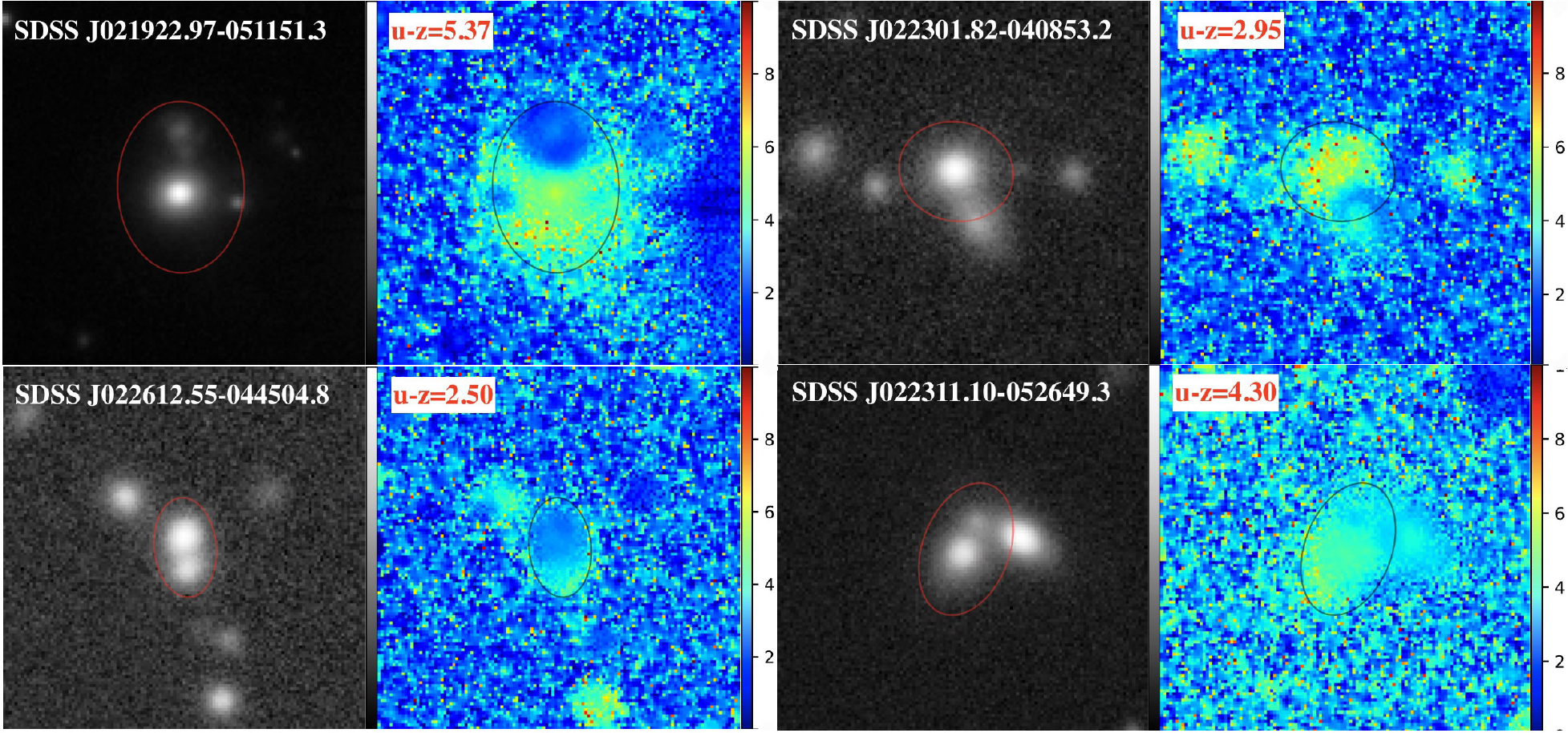}
    \caption{The $z$ band images and $u-z$ color maps of 4 galaxies in  Category I. Each panel covers a $20\arcsec\times20\arcsec$ area. Red (in $z$ band images) and black (in $u-z$ color maps) ellipses indicate $z$ band Petrosian radii of the galaxies.}
    \label{fig:ap_category1}
\end{figure*}

\begin{figure*}
    \epsscale{1.15}
    \plotone{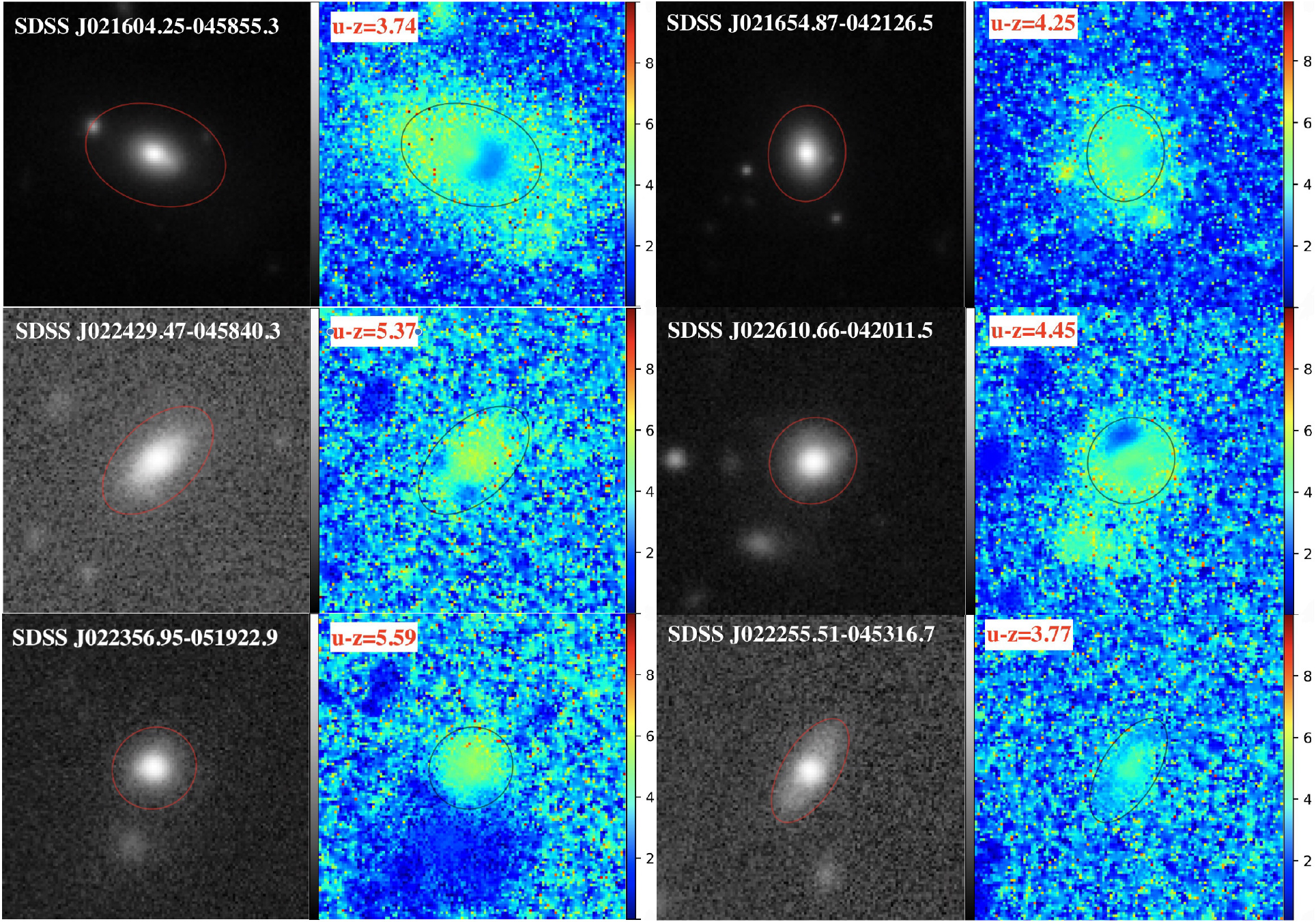}
    \caption{Same as Figure \ref{fig:ap_category1}, but for 6 galaxies in Category II.}
    \label{fig:ap_category2}
\end{figure*}

\begin{figure}
    \epsscale{1.15}
    \plotone{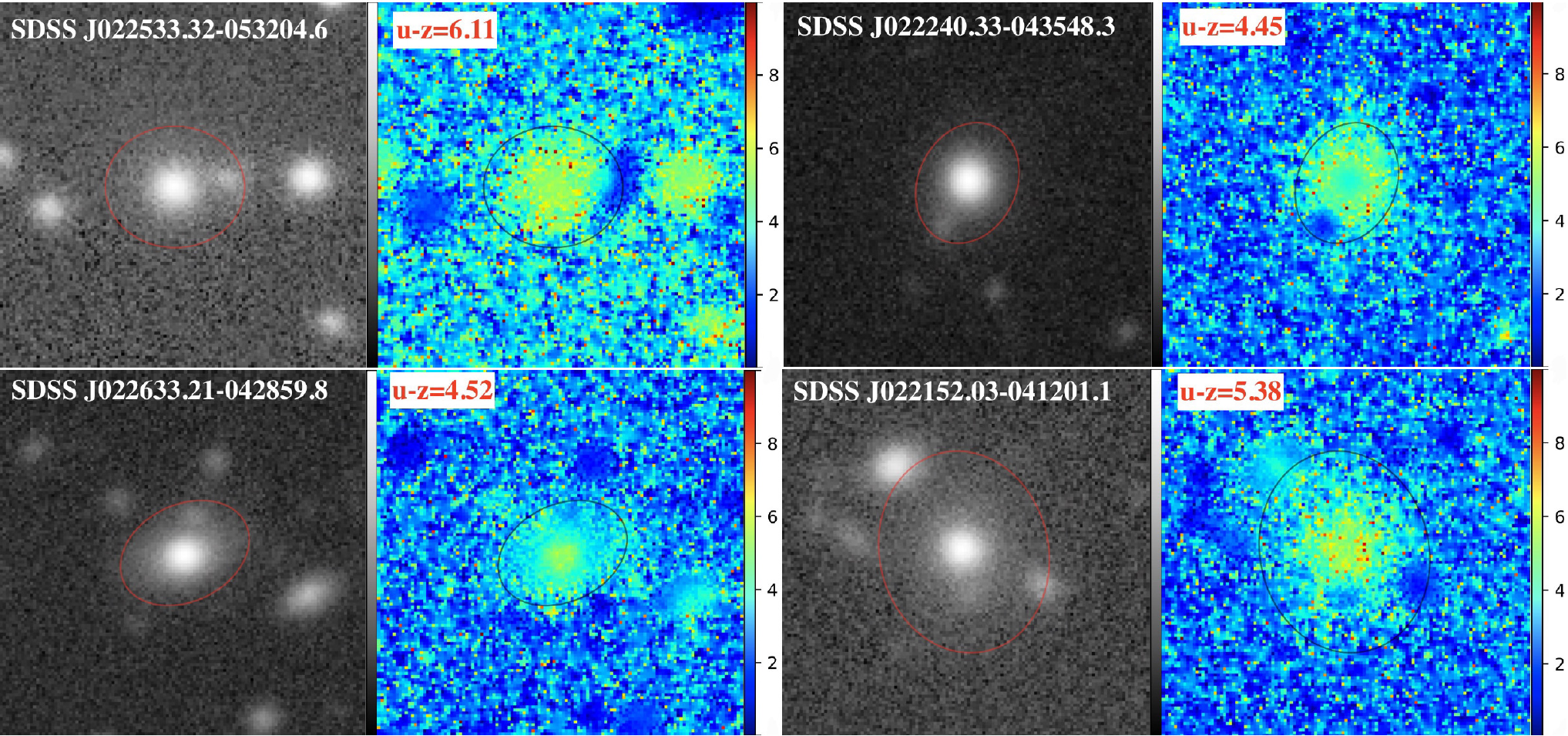}
    \caption{Same as Figure \ref{fig:ap_category1}, but for 4 galaxies in Category I-II.}
    \label{fig:ap_category12}
\end{figure}

In Figure \ref{fig:ap_category1}, \ref{fig:ap_category2} and \ref{fig:ap_category12}, we show the $z$ band images and $u-z$ color maps for 4 galaxies in Category I, 6 galaxies in Category II and 4 galaxies in Category I-II, respectively. The images of galaxies in Category III have been already shown in Figure \ref{fig:f4}.

\clearpage

\bibliographystyle{aasjournal}
\bibliography{mybib}{}

\end{CJK*}
\end{document}